\documentclass[aip,reprint]{revtex4-1}
\usepackage{graphicx}
\usepackage{epsfig}
\usepackage{color}
\usepackage{epstopdf}
\usepackage{amsmath}
\usepackage{hyperref} 
\usepackage{bm}

\newcommand{\deriv}[2]{\frac{\partial #1}{\partial #2}}

\newcommand{\grl}{    {Geophys. Res. Lett.}}
\newcommand{\jgr}{    {J. Geophys. Res.}}
\newcommand{\ssr}{    {Space Sci. Rev.}}

\newcommand{\physrep}{ {Phys. Report}}

\begin{document}


\title{Ion motion in a polarized current sheet}




\author{E. Tsai}
\affiliation{
Institute of Geophysics and Planetary Physics, UCLA, Los Angeles, California, USA.
}%
\email{ethantsai@ucla.edu}

\author{A. V. Artemyev}
 \altaffiliation[Also at ]{Space Research Institute, RAS, Moscow, Russia} 
\affiliation{
Institute of Geophysics and Planetary Physics, UCLA, Los Angeles, California, USA.
}%

\author{V. Angelopoulos}
\affiliation{
Institute of Geophysics and Planetary Physics, UCLA, Los Angeles, California, USA.
}%

\date{\today}

\begin{abstract}
We consider the effects of a polarization electric field on transient ion motion in a thin current sheet. Using adiabatic invariants, we analytically describe a variety of ion trajectories in current sheet configurations which include a local minimum or maximum of the scalar potential in the central region. Ions in the current sheet can either be trapped or ejected more efficiently than in an unpolarized current sheet, depending on the sign and magnitude of the polarization electric field. We derive an expression for the relative phase space volume filled by transient particles as a function of the electric field amplitude. This expression allows us to estimate the dependence of transient particle and current densities on the electric field. We discuss  applicability of these results for current sheets observed in planetary magnetospheres.
\end{abstract}


\maketitle 

\section{Introduction}

Spacecraft observations in Earth's magnetotail \cite{Jackman14,Petrukovich15:ssr} and laboratory simulations \cite{Yamada00:cs, Frank11, Olson16} demonstrate that the current sheet is a key element of plasma systems with high plasma pressure. But many details of current sheet configurations are unknown; specifically, current density generation in very thin current sheets has not yet been well described theoretically. Models of such current sheets include specific ion velocity distributions \cite{Zelenyi00, Sitnov00, Sitnov06, Zelenyi11PPR} shaped by particles residing within the current sheet for a finite time interval  \cite{Speiser65, BZ89}. Although these models successfully describe many properties of current sheets observed in Earth's magnetotail\cite{Zhou09, Artemyev&Zelenyi13, Zhou16}, significant discrepancies between  model predictions and observations still exist \cite{Runov06, Israelevich08, Artemyev09:angeo}. One of these discrepancies is the observed dominance of currents carried by cold electrons. To mitigate this, modern current sheet models include electrostatic fields (Hall electric fields\cite{SB02, YL04, Zelenyi11GRL}) generated by decoupling of ion motions from electron motions within the strong magnetic field gradients of thin current sheets \cite{Schindler12}. These fields should induce ${\bf E}\times {\bf B}$ drifts that redistribute currents between ions and electrons and make electron currents stronger, as observed by spacecraft.

\begin{figure*}
\includegraphics[width=35pc]{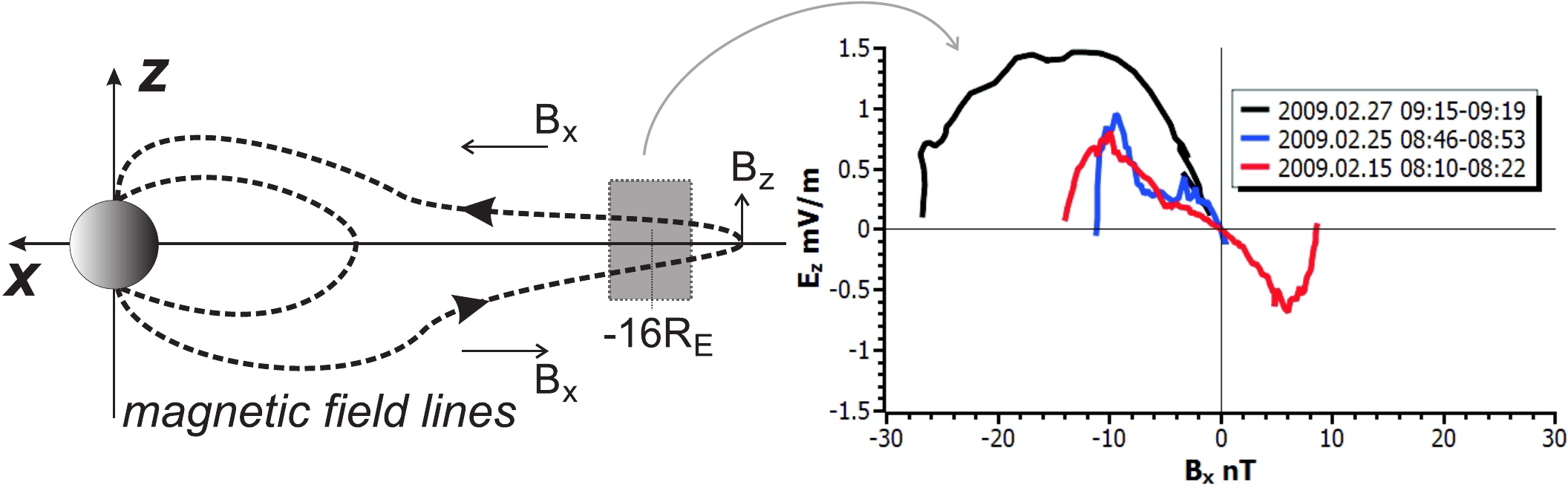}
\caption{Left panel: a schematic view of Earth's magnetotail. Right panel: electric field $E_z(B_x)$ profiles for three events when (Time History of Events and Macroscale Interactions during Substorms) THEMIS C probed the magnetotail current sheets around $x\sim -16R_E$ downtail ($R_E\approx 6400$ km is
Earth's radius). For two events, the spacecraft crossed only half the current sheet with $B_x<0$. Details of $E_z$ calculations are given in Appendix A.}
\label{scheme}
\end{figure*}

And indeed, there is some observational \cite{Wygant05, Artemyev11:jgr, Vasko14:angeo} and modeling \cite{Greco07, Schindler&Hesse08, Lin14:hybrid_code, Lu16:assymetry} evidence of such Hall electric fields in thin current sheets. In the coordinate system shown in Fig. \ref{scheme} (right panel), the main Hall electric field component is directed along the normal to the current sheet's neutral plane, $z=0$. Because direct measurements of the $E_z$ field in the magnetotail current sheet are significantly complicated by spacecraft orientation, most measurements were performed for particular tilted  current sheets with the neutral plane $y\approx 0$ (see Refs. \onlinecite{Wygant05, Vasko14:angeo}). In the absence of direct $E_z$ measurements, this electric field component can also be derived from accurate analysis of electron motion in the current sheet. Using such analysis, we reconstruct the profile of $E_z$ across the current sheet observed in Earth's magnetotail (see Appendix for details). Figure \ref{scheme} (left panel) demonstrates that the electric field amplitude can reach a few mV/m. Such strong electric fields certainly result in plasma drifts $\sim E_z B_x/{\bf B}^2$, which can increase the electron velocity and decrease the ion velocity. Therefore, for magnetized ions drifting in current sheet electromagnetic fields, the observed electric field can be responsible for the ion current density reduction.

However, unmagnetized transient ions make up a significant portion of the ion population in thin current sheets. Because these ions exhibit complex motion that includes meandering around the current sheet's neutral plane \cite{Speiser65, Sonnerup71, BZ86}, drift theory cannot describe their motion in an electric field $E_z$. For a typical magnetotail current sheet configuration, however, we can separate transient ion motion into fast and slow motions, providing an opportunity to apply adiabatic theory \cite{Zelenyi13:UFN, Zhou16}. In this study, we adopt adiabatic theory methods to describe transient ion dynamics in the current sheet with electric field $E_z$ (see also Ref. \onlinecite{Dolgonosov10}). Our main objective is to determine the dependence of transient ion currents on the electric field $E_z$. We consider individual particle motion and use adiabatic invariants to model the main characteristics of this motion. Next, we demonstrate how the boundary between transient and trapped ions in the phase space depends on the electric field $E_z$. Using this knowledge, we determine the dependence of transient ion current density on $E_z$ in thin current sheets. Further applications of our results are discussed in the final section.

\section{Basic equations}
Current sheet magnetic and electric fields shown in Fig. \ref{scheme} can be approximated by a simple analytic model
\begin{align}
{\bf B} = B_0 \frac{z}{L} {\bf e}_x + B_z {\bf e}_z, \quad {\bf E} = - E_0 \frac{z}{L} {\bf e}_z
\label{analyticalModel}
\end{align}
where $L$ is the current sheet thickness. Model (\ref{analyticalModel}) captures the essential features of the current sheet configuration and allows us to carry out appropriate analysis of the particle trajectories: $B_x$ changes sign at the neutral plane ($z = 0$), $B_z$ is constant across the sheet, and a polarizing electric field $E_z$ is directed either outward ($E_0>0$) or inward ($E_0<0$) relative to the neutral plane. Model (\ref{analyticalModel}) approximates only the central current sheet region where a strong current density supports magnetic field gradients. At the current sheet boundary $z=z_{max}$, the magnetic field $B_x$ approaches constant values $\pm B_0(z_{\max}/L)$.

Electromagnetic fields given by Eq. (\ref{analyticalModel}) correspond to the following vector and scalar potentials
\begin{align}
	{\bf A} = \Big( B_z x -  \frac{B_0 z^2}{2L} \Big){\bf e}_y, \quad \varphi = \frac{E_0 z^2}{2L}
	\label{potentials}
\end{align}
where the Coulomb gauge is used for the vector potential, $\nabla\cdot{\bf A}=0$, and the two magnetic field components are described by $\partial A_y/\partial y=0$.
The Hamiltonian for a particle with charge $q$ and mass $m$ moving in the given electromagnetic fields then has the form
\begin{align*}
\mathcal H = \frac{1}{{2m}}\Big( {p_x^2  + p_z^2 } \Big) + \frac{1}{{2m}}\Big( {p_y  - qA_y (x,z)} \Big)^2  + q\varphi (z)
\end{align*}
Here ${\bf p}=p_x{\bf e}_x+p_y{\bf e}_y+p_z{\bf e}_z$ is the generalized particle momentum. Taking into account the conservation of $p_y$ (due to $\partial \mathcal H/\partial y=0$) and using Eq. (\ref{potentials}), we obtain
\begin{align}
  \mathcal H = \frac{{p_x^2 + p_z^2}}{2m} + \frac{1}{2m} \Bigg(q B_z x - \frac{q B_0 z^2}{2L}\Bigg)^2 + \frac{q E_0 z^2}{2 L}
  \label{dimensionalHamiltonian}
\end{align}
We introduce dimensionless variables $x/l\to x$, $z/l\to z$, $p_x/p_0\to p_x$, $p_z/p_0\to p_z$; dimensionless Hamiltonian $\mathcal{H} \to \mathcal H m/p_0^2$; and dimensionless time $tp_0/ml\to t$, where $p_0$ is a typical particle momentum (i.e., normalized value of $\mathcal{H}$ equals to $1/2$), $l=\sqrt{L\rho_0}$, and $\rho_0=qB_0/p_0$ is the particle Larmor radius. The new dimensionless Hamiltonian takes the form
\begin{align}
    \mathcal H = \frac{1}{2} \Big(p_x^2 + p_z^2\Big) + \frac{1}{2}\Big(\kappa x - \frac{1}{2} z^2\Big)^2 + \frac{1}{2} \alpha z^2
    \label{dimensionlessHamiltonian}
\end{align}
where
\begin{align}
  \kappa = \frac{B_z}{B_0}\sqrt{\frac{L}{\rho_0}}  = \sqrt{\frac{R_c}{\rho_z}}, \quad \alpha = \frac{E_0 m}{B_0 p_0} = \frac{m v_d}{p_0}
  \label{kappaalpha}
\end{align}
The $\alpha$ parameter here, in Eq. (\ref{kappaalpha}), is proportional to ${\rm sign}(q)$; since we focus mainly on ion motion in this study, $q$ remains positive. However, obtained results are applicable to electrons as well by merely changing $\alpha \to -\alpha$.

Equation (\ref{kappaalpha}) shows that $\kappa$ represents the ratio of the magnetic field curvature radius $R_c=B_zL/B_0$ to the particle Larmor radius in the neutral plane $\rho_z=\rho_0 B_0/B_z$. $\alpha$ represents the ratio of the ${\bf E} \times {\bf B}$ drift velocity $v_d$ to the typical particle velocity $p_0/m$; more specifically, $\alpha$ is a parameter that denotes the amplitude of the electric field. The Hamiltonian (\ref{dimensionlessHamiltonian}) depends only on these two parameters: $\kappa$ and $\alpha$. There are three regimes of particle motion for various $\kappa$ values (see Refs. \onlinecite{Chen92,Horton97,Delcourt&Belmont98}): $\kappa\gg 1$ corresponds to well magnetized particles (the magnetic field inhomogeneous scale, $R_c$, is much larger than the particle Larmor radius) described by the theory of drift motion \cite{bookNorthrop63, bookSivukhin65}; $\kappa \sim 1$ corresponds to chaotic particle motion\cite{Chen86,Delcourt94:scattering,Delcourt04,Shustov15}; $\kappa\ll 1$ corresponds to separation of particle motion timescales in the $x$ and $z$ directions. This latter regime of motion is typical for hot ions in planetary magnetotail current sheets \cite{Zelenyi13:UFN}. Particles with small $\kappa$ move along transient orbits and carry significant currents \cite{Pritchett92, Burkhart92TCS, Zelenyi11PPR, Sitnov&Merkin16}. Ions in this region are the focus of our investigation. We begin with the well-explored case of zero electric field, $\alpha=0$, and then consider a more general one $\alpha\ne0$.

From Eq. (\ref{dimensionlessHamiltonian}), the Hamiltonian equations of motion can be written as

\begin{align*}
  \dot z = \deriv{\mathcal H}{p_z} = p_z, \quad \dot p_z = - \deriv{\mathcal H}{z} = z \bigg( \kappa x - \frac{z^2}{2}\bigg)   \label{EoM}
\end{align*}
\begin{align}
  \kappa \dot x = \kappa \deriv{\mathcal H}{p_x} = \kappa p_x, \quad \dot p_x = - \kappa \deriv{\mathcal H}{\kappa x} = - \kappa \bigg( \kappa x - \frac{z^2}{2}\bigg)
\end{align}
Note that from the Lorentz force, $\dot{p}_x = \kappa \dot y \rightarrow p_x = \kappa y + const$. In other words, the particle momentum $p_x$ is linearly proportional to the particle coordinate $\kappa y$. We then consider projections of particle trajectories to the phase plane $(p_x,\kappa x)$, which coincides with the neutral plane $z=0$. Solutions of Eqs. (\ref{EoM}) represent trajectories in 4D phase space $(\kappa x, p_x, z, p_z)$. Because of energy conservation ($\partial \mathcal H/\partial t=0$), however, particle trajectories are confined in 3D space $(\kappa x, p_x, z)$, which then coincides with $(\kappa x, \kappa y, z)$.

\begin{figure}
\includegraphics[width=20pc]{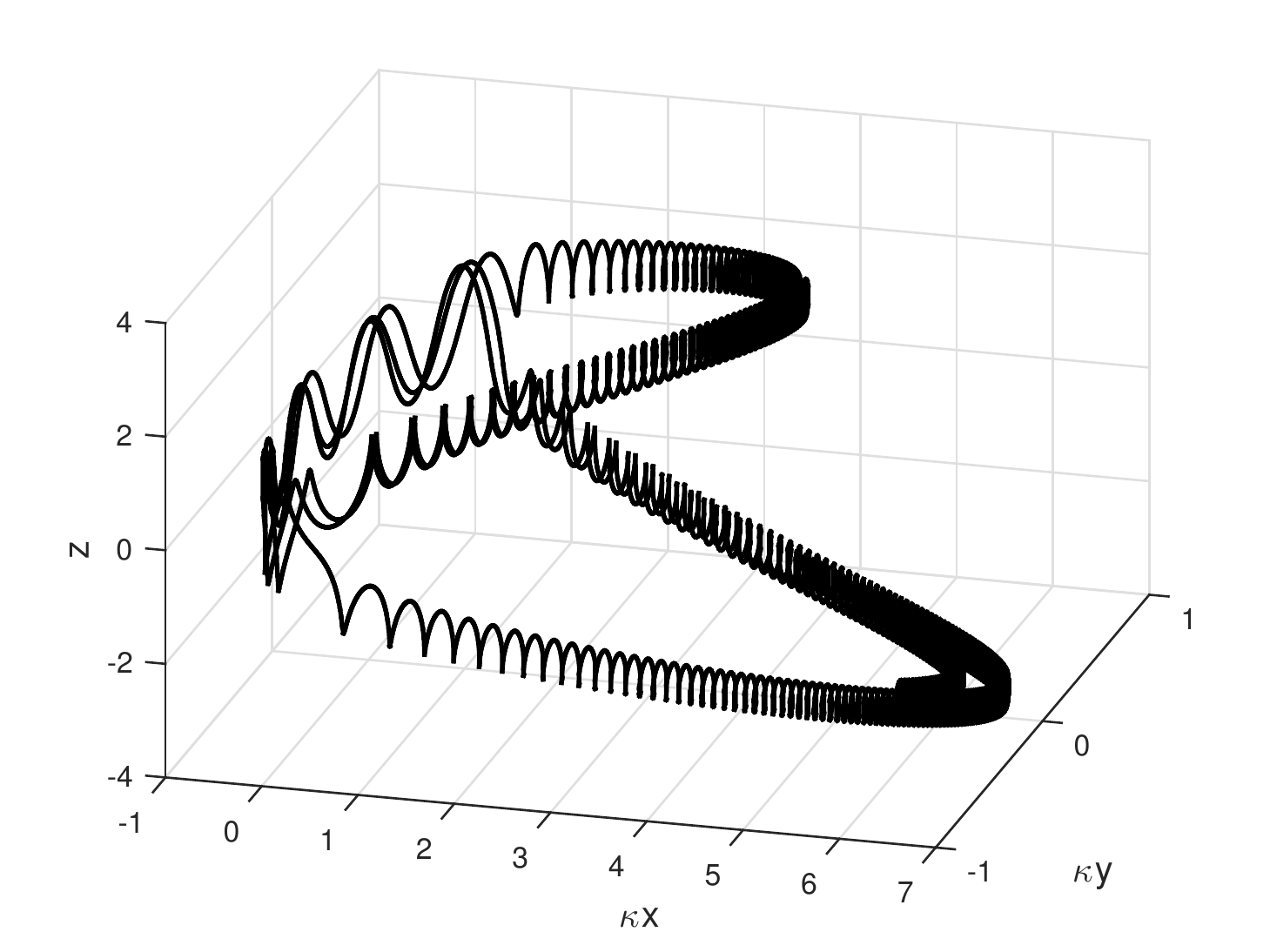}
\caption{3D Trajectory of particle}
\label{trajectory}
\end{figure}

A particle trajectory obtained from numerical integration of Eqs. (\ref{EoM}) is shown in Fig. \ref{trajectory}. For $\kappa x > 0$, the charged particle moves along the magnetic field lines either toward or away from the neutral plane $z=0$; these two diverging particle trajectory elements can be seen branching out above and below the neutral plane. At some coordinate above/below $z=0$, the particle traveling away from the neutral plane turns around and starts moving back towards the neutral plane. It reaches the neutral plane at around $\kappa x = 0$; then it makes a half rotation in the $z=0$ plane and  escapes the neutral plane. When it escapes that plane, the particle can go to either positive $z$ or negative $z$. In Fig. \ref{trajectory}, the particle has been traced for a sufficiently long time to move along both types of trajectories.

\begin{figure}
\includegraphics[width=20pc]{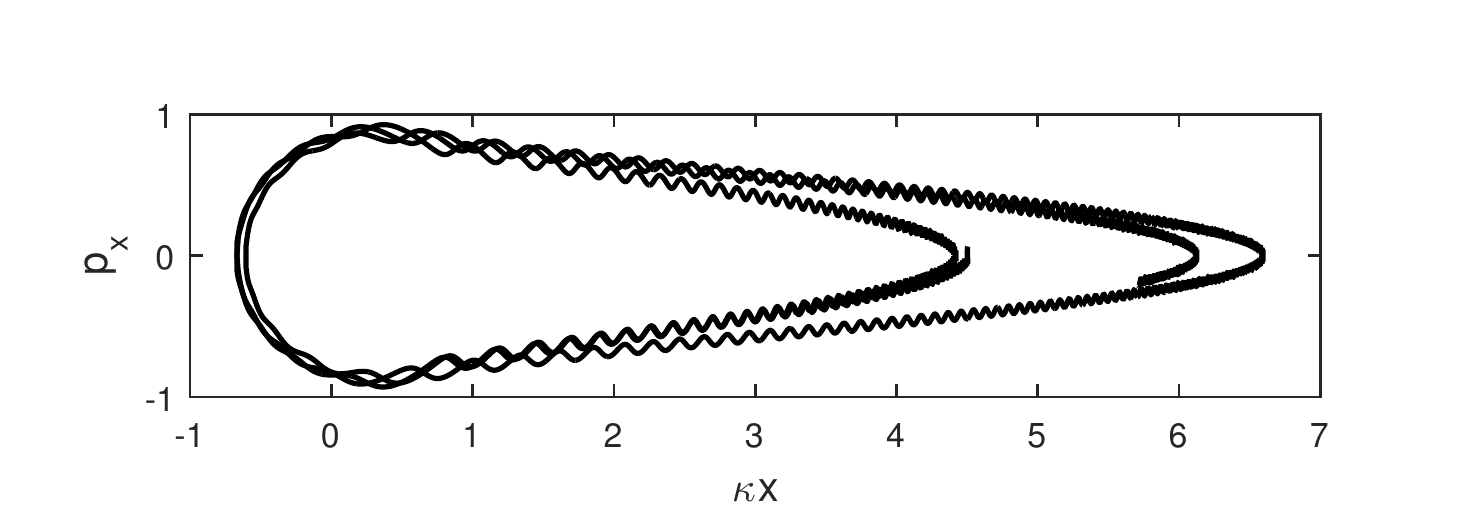}
\caption{Trajectory projection to the $(\kappa x,p_x)$ plane for $\kappa = 0.1$, $\alpha = 0$.}
\label{slowtraj}
\end{figure}

\begin{figure}
\includegraphics[width=20pc]{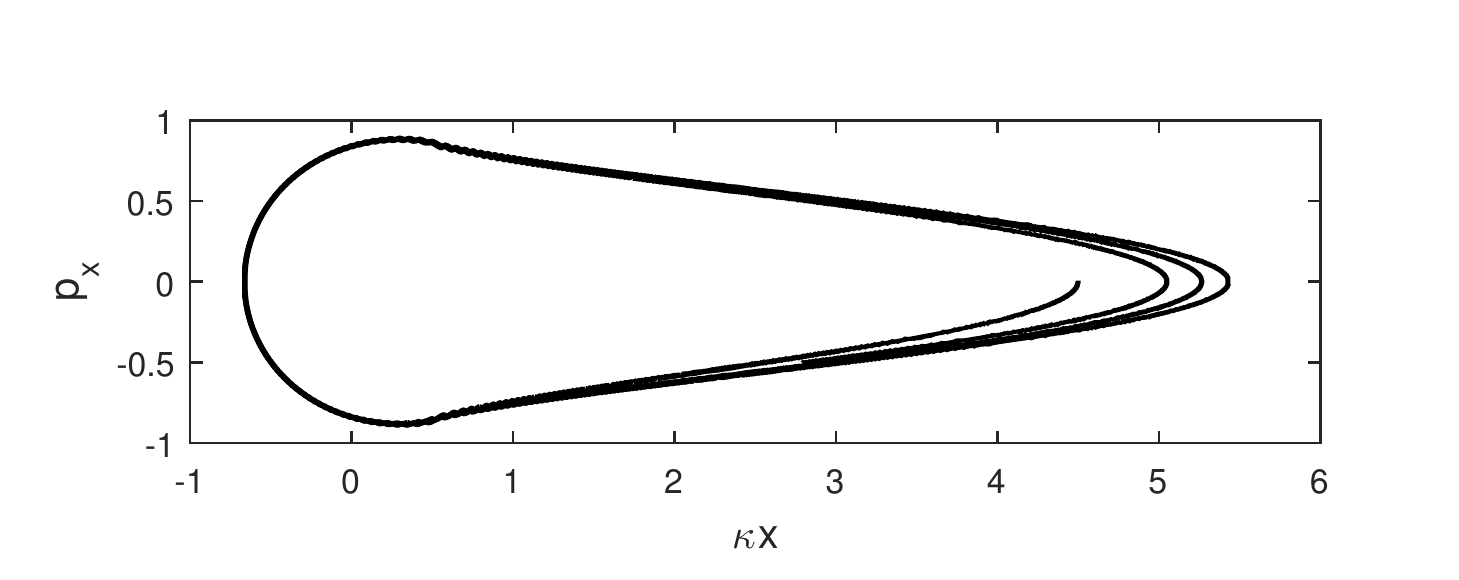}
\caption{Trajectory projection to the $(\kappa x,p_x)$plane for $\kappa = 0.01$, $\alpha = 0$.}
\label{slowtrajsmooth}
\end{figure}

The projection of this trajectory onto the $(\kappa x, p_x)$ plane is shown in Fig. \ref{slowtraj}. Note that the particle does not move along a closed trajectory in the $(\kappa x, p_x)$ plane; rather, it drifts from one nearly closed trajectory to another one because of particle scattering around the neutral plane $z=0$. In a sufficiently long time interval, a single particle's trajectory will fill nearly the entire  $(\kappa x, p_x)$ plane\citep{BZ89, Zelenyi13:UFN}. The rate of particle scattering is $\sim\kappa$. When Fig. \ref{slowtraj} and Fig. \ref{slowtrajsmooth} are compared, a smaller $\kappa$ can be seen to correspond to weaker particle scattering.

According to equations (\ref{EoM}), the variation rate of $(z, p_z)$ is about $O(1)$ (i.e., it does not depend on $\kappa$), whereas the variation rate of $(\kappa x, p_x)$ is about $O(\kappa)$. Thus, $\kappa x$ and $p_x$ change at a much slower rate than $(z, p_z)$, which allows us to separate variables into slow $(\kappa x, p_x)$ and fast $(z, p_z)$. And, indeed, the $z$-oscillations in the trajectory shown in Fig. \ref{trajectory} are much faster than particle rotation in the $(\kappa x, p_x)$ plane. This separation of particle motion timescales permits application of adiabatic theory to the Hamiltonian system (\ref{dimensionlessHamiltonian}).

\section{Adiabatic invariant}
\begin{figure}
\includegraphics[width=20pc]{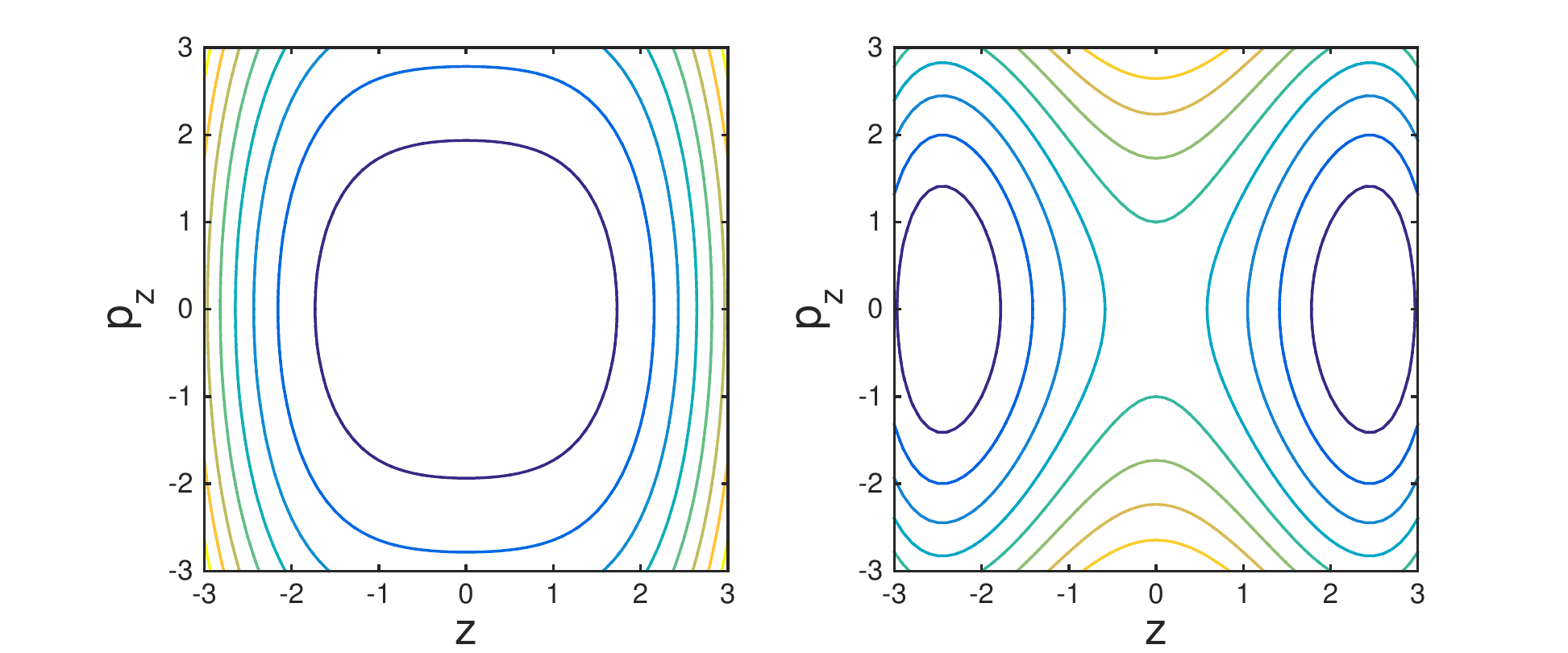}
\caption{Level lines of the fast Hamiltonian that correspond to the two different trajectories of the ion in the $z$, $p_z$ plane (different colors for different $h_z$ values).}
\label{fig:fastHamiltonian}
\end{figure}

\begin{figure}
\includegraphics[width=20pc]{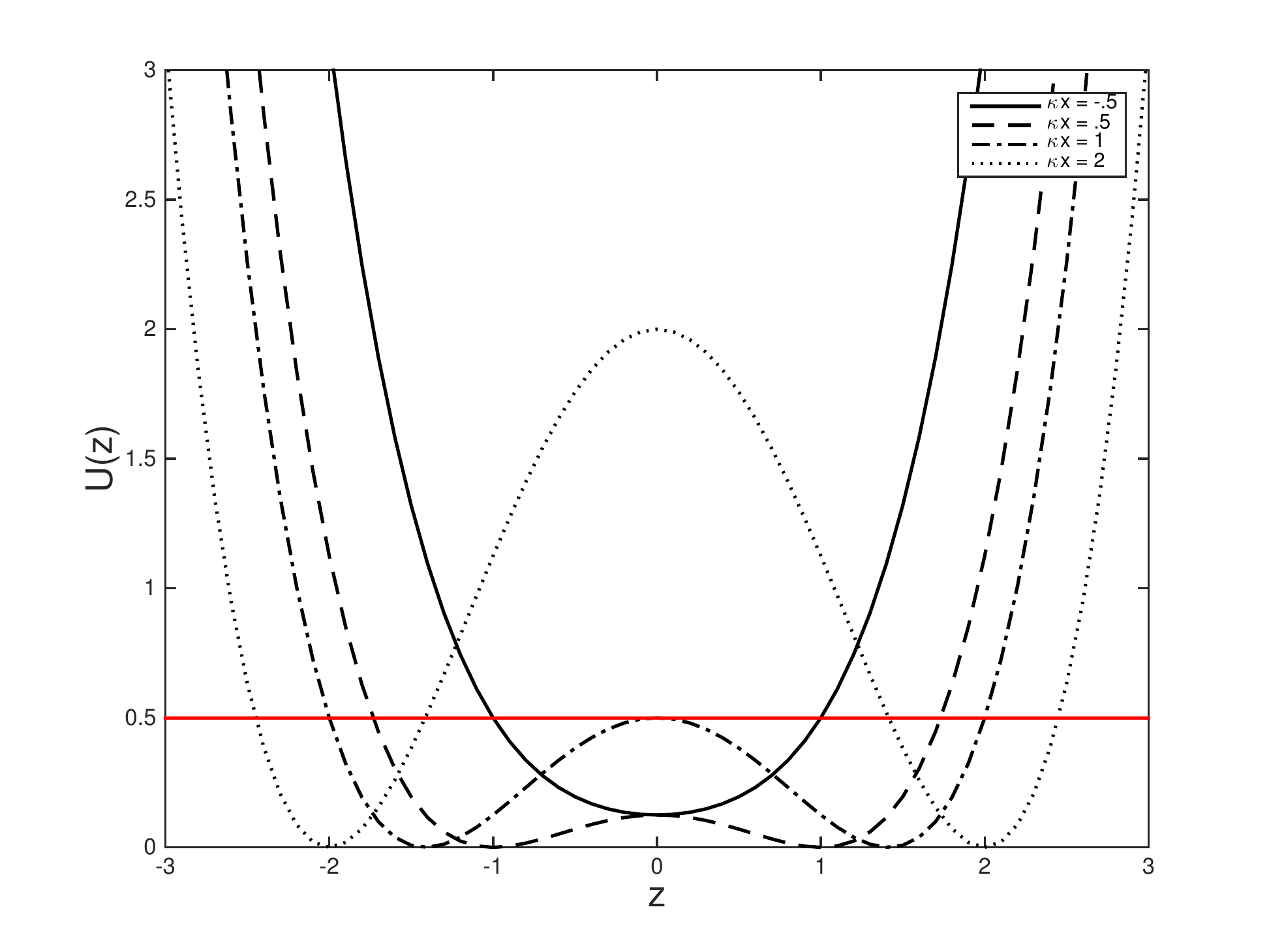}
\caption{Potential energy for various values of $\kappa x$. The solid red line denotes the energy $U$ level corresponding to change of particle motion from oscillations in one of two small potential wells to oscillations above these wells.}
\label{fastPotEnergy}
\end{figure}

Separation of $(\kappa x, p_x)$ and $(z, p_z)$ variation timescales allows us to consider system dynamics in the $(z, p_z)$ plane with frozen slow variables $(\kappa x, p_x)$. For simplicity, we begin with $\alpha = 0$ and write the Hamiltonian for the fast motion
\begin{align}
  h_z = \mathcal H - \frac{1}{2} p_x^2 =
  \frac{1}{2} p_z^2 + \frac{1}{2} \Big(\kappa x - \frac{1}{2} z^2\Big)^2
  \label{fastHamiltonian}
\end{align}
where $\kappa x$ is a constant parameter. For each fixed value of $\mathcal H$ and $\kappa x$, a value of $h_z$ determines the trajectory in the $(z, p_z)$ plane. Figure \ref{fig:fastHamiltonian} shows two sets of particle trajectories plotted for $\kappa x<0$ and $\kappa x>0$. All trajectories are closed; from the Hamiltonian (\ref{fastHamiltonian}), we see that the particles oscillate within the effective potential well
\begin{align}
  U = \frac{1}{2}\Big(\kappa x - \frac{1}{2}z^2\Big)^2
\end{align}
Figure \ref{fastPotEnergy} demonstrates how a system with two symmetric potential wells ($\kappa x>1$) transforms to a system with a single potential well ($\kappa x < 0$). Comparing Fig. \ref{trajectory} with Fig. \ref{fastPotEnergy}, we note that particle oscillations within one of two potential wells correspond to particle motion below or above the neutral plane $z=0$, whereas particle oscillations within the single potential well correspond to particle oscillations within the neutral plane $z=0$.

\begin{figure}
\includegraphics[width=20pc]{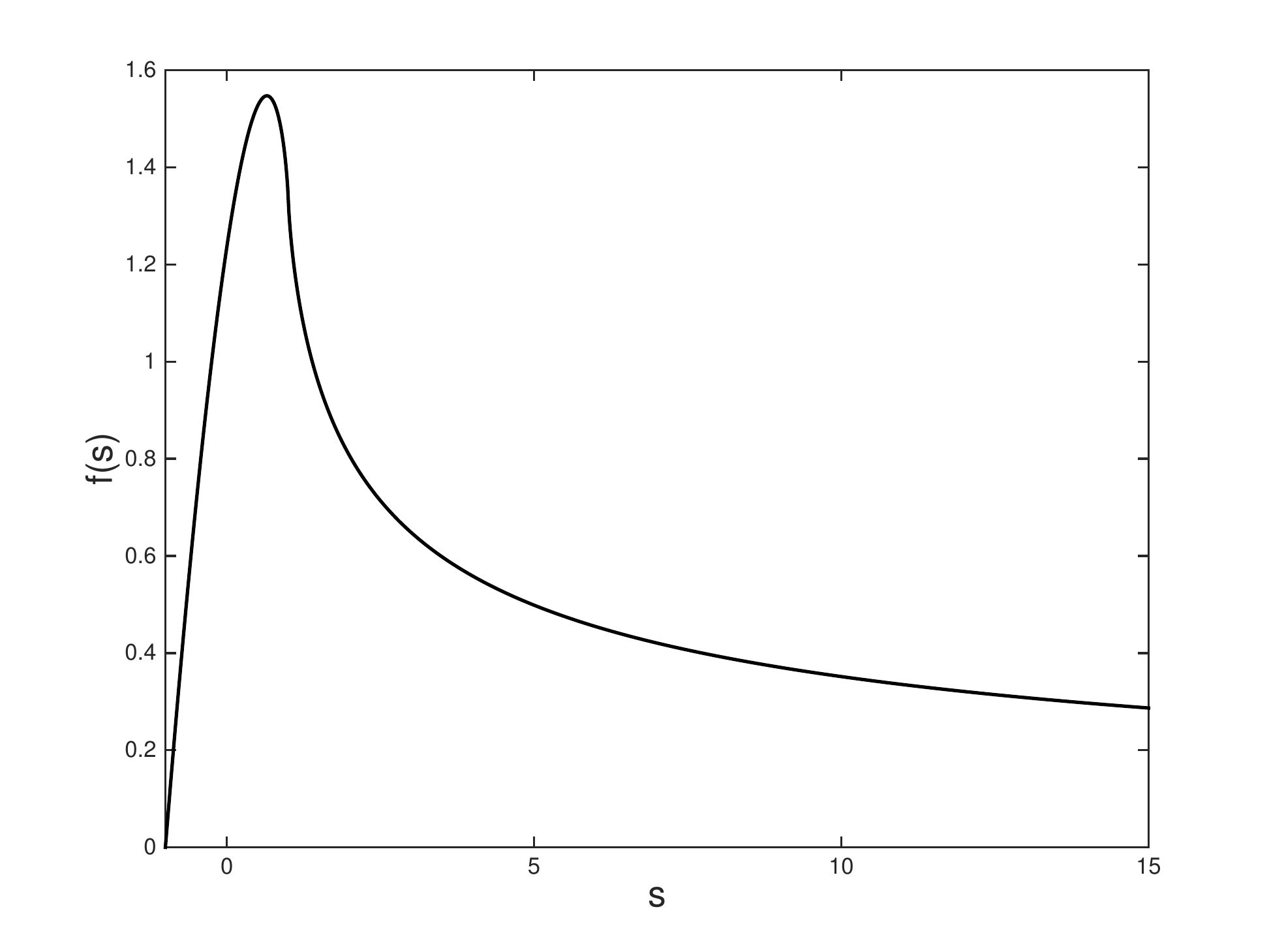}
\caption{A plot of $f(s)$.}
\label{fig:fs}
\end{figure}

The periodicity of particle oscillations in the fast variable plane $(z,p_z)$ allows us to introduce the adiabatic invariant of motion\cite{bookLL:mech}
\begin{align}
  I_z = \oint p_z dz
  \label{genAdiabaticInvariant}
\end{align}
Definition (\ref{genAdiabaticInvariant}) assumes that the adiabatic invariant should be calculated for fixed slow variables ($\kappa x, p_x = const$), and the adiabatic invariant $I_z$ equals the area enclosed by the trajectory in the $(z,p_z)$ plane. Using Eq. (\ref{fastHamiltonian}), we rewrite Eq. (\ref{genAdiabaticInvariant}) as
\begin{align}
  I_z = 2 (2h_z)^{3/4} \int_{\zeta_1}^{\zeta_2} \sqrt{1 - \Big( s - \frac{1}{2} \zeta\Big)^2} d\zeta
  \label{adiabaticInvariantAlmostSimplified}
\end{align}
where
\begin{align}
  s = \frac{\kappa x}{\sqrt{2h_z}}, \quad \zeta = \frac{z}{(2h_z)^{1/4}}
  \label{syvariables}
\end{align}
and $\zeta_{1,2} = \pm \sqrt{2} \sqrt{s \pm 1}$. There are two cases: for $s>1$ (i.e., $\kappa x>\sqrt{2h_z}$), there are four roots where $p_z=0$ caused by two symmetrical trajectories in the $(z,p_z)$ plane, whereas for $s<1$ (i.e., $\kappa x<\sqrt{2h_z}$), there are just two roots (see Fig. \ref{fig:fastHamiltonian}). We introduce the function
\begin{align}
	f(s) = \int_{\zeta_{\min}}^{\zeta_{\max}} \sqrt{1 - \Big( s - \frac{1}{2} \zeta\Big)^2} d\zeta
	\label{fs}
\end{align}
and write the final equation for $I_z=2(2h_z)^{3/4}f(s)$. The limits of integration in Eq. (\ref{fs}) are $\zeta_{\max}=\sqrt{2} \sqrt{s + 1}$,  $\zeta_{\min}=\sqrt{2} \sqrt{s - 1}$ for $s>1$, and $\zeta_{\min}=0$ for $s\leq 1$. Figure (\ref{fig:fs}) shows the  $f(s)$ profile.

\begin{figure}
\includegraphics[width=20pc]{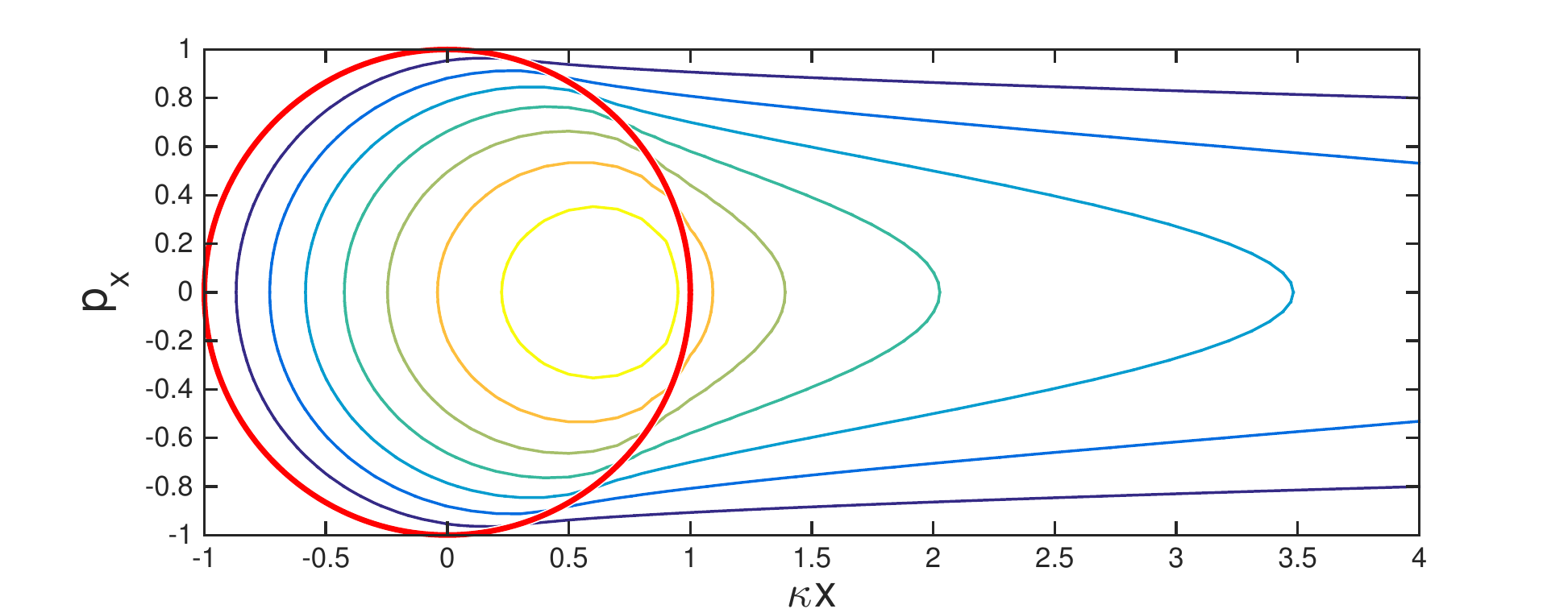}
\caption{Level lines of the adiabatic invariant that correspond to trajectories in the $(\kappa x, p_x)$ plane. The solid red line denotes the curve where $\kappa x=\sqrt{2h_z}$ for $z=0$, $p_z=0$.}
\label{adiabaticInvariantTraj}
\end{figure}

\begin{figure}
\includegraphics[width=20pc]{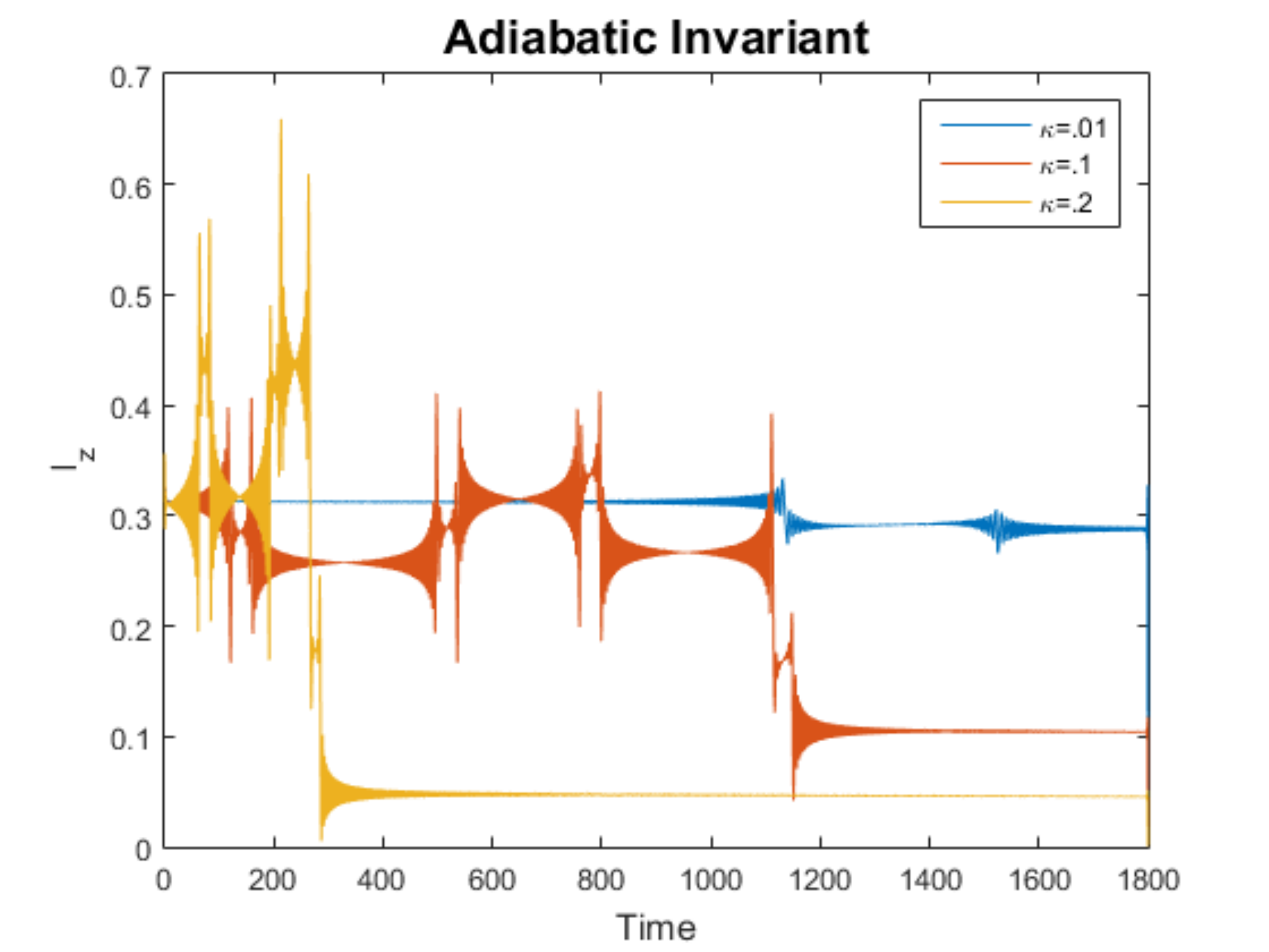}
\caption{The adiabatic invariant over time for various $\kappa$. The jumps are on the order of $\sim\kappa$}
\label{adiabaticInvariantJumps}
\end{figure}

The invariant $I_z$ depends on $(\kappa x, p_x)$; thus, conservation of $I_z$ defines the set of particle trajectories in the slow variable plane. Figure (\ref{adiabaticInvariantTraj}) shows that the contour of $I_z(\kappa x, p_x)=const$ looks very similar to projections shown in Figs. \ref{slowtraj}, \ref{slowtrajsmooth}. $I_z$ is well conserved (with an accuracy about $\kappa$, see Ref. \onlinecite{Neishtadt86, Cary86}) when $(\kappa x, p_x)$ and $(z, p_z)$ timescales are separated. However, when particles change their motion from oscillations within one of two potential wells to oscillations within a single well, invariant $I_z$ experiences random jumps with amplitude $\kappa$. To demonstrate this effect, we calculate $I_z$ given by Eq. (\ref{adiabaticInvariantAlmostSimplified}) along numerically integrated particle trajectories (see Fig. \ref{adiabaticInvariantJumps}). Each jump of the adiabatic invariant shown in this figure corresponds to a change in the type of particle motion (see more details about $I_z$ destruction in Ref. \onlinecite{BZ89,Zelenyi13:UFN}). The amplitude of $I_z$ jumps decreases when $\kappa$ decreases.

\begin{figure}
\includegraphics[width=15pc]{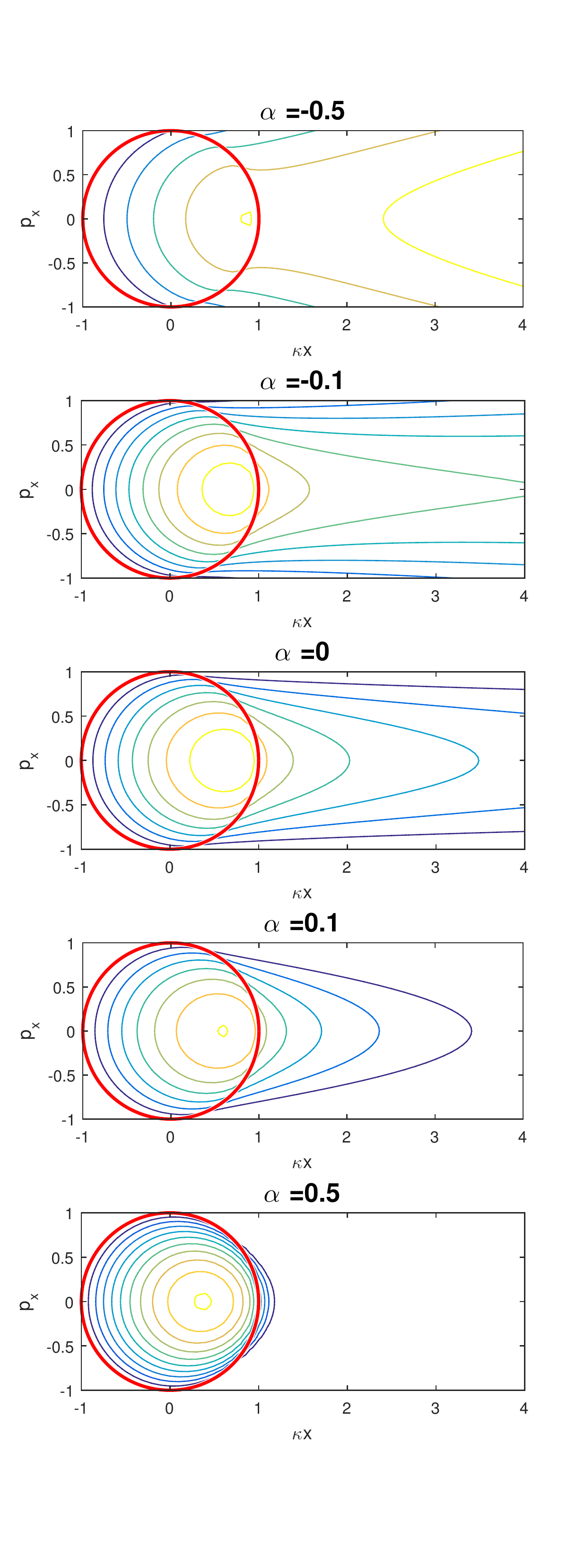}
\caption{Same as in Fig. \ref{adiabaticInvariantTraj}, but for various $\alpha$ values. }
\label{comparetrajalpha}
\end{figure}

As seen in the Hamiltonian (\ref{dimensionlessHamiltonian}), an electric field only adds a term $\frac{1}{2} \alpha z^2$, where $\alpha$ defines the electric field amplitude. To introduce this term into the adiabatic invariant, we rewrite the Hamiltonian (\ref{dimensionlessHamiltonian}) as
\begin{align}
   	\mathcal H = \frac{1}{2} \Big(p_x^2 + p_z^2\Big) + \frac{1}{2}\Big(\kappa x - \alpha - \frac{z^2}{2} \Big)^2 + \alpha \kappa x - \frac{\alpha^2}{2}
   	\label{newHamiltonian}
\end{align}
We use a tilde to denote the new term for $\alpha \neq 0$. The new fast Hamiltonian $\tilde{h}_z$ has the form
\begin{align}
  \tilde{h}_z = \mathcal H - \frac{1}{2} p_x^2 - \alpha \kappa x + \frac{1}{2} \alpha ^2 = h_z - \alpha \kappa x + \frac{1}{2} \alpha ^2
  \label{newhz}
\end{align}
This new Hamiltonian $\tilde {h}_z$ depends on $z$ in the same way as $h_z$ does in Eq. (\ref{fastHamiltonian}). Thus, we can introduce a new parameter $\kappa \tilde{x} = \kappa x - \alpha$ and redefine the $s$ and $\zeta$ variables in Eq. (\ref{syvariables}) as follows
\begin{align}
  \tilde s &= \frac{\kappa \tilde{x}}{\sqrt{2\tilde{h}_z}} = \frac{\kappa x - \alpha}{\sqrt{2h_z - 2\alpha \kappa x + \alpha^2}} \notag\\
  \tilde \zeta &= \frac{z}{(2\tilde{h}_z)^{1/4}} = \frac{z}{(2h_z - 2\alpha \kappa x + \alpha^2)^{1/4}}
  \label{syvariables_new}
\end{align}
Using these new variables allows the adiabatic invariant  (\ref{adiabaticInvariantAlmostSimplified}) to be recalculated as $I_z = 2 (2\tilde{h}_z)^{3/4} \cdot f(\tilde s)$. Note that the electric field does not change the form of the invariant $I_z$; rather, $I_z$ now depends on $(\kappa x, p_x)$ through $\tilde{s}$. Figure \ref{comparetrajalpha} shows that for $\alpha > 0$, trajectories become shorter; in contrast, for $\alpha<0$, more ions escape the vicinity of $\kappa x\sim 0$ and move to larger $\kappa x$.

\begin{figure}
\includegraphics[width=20pc]{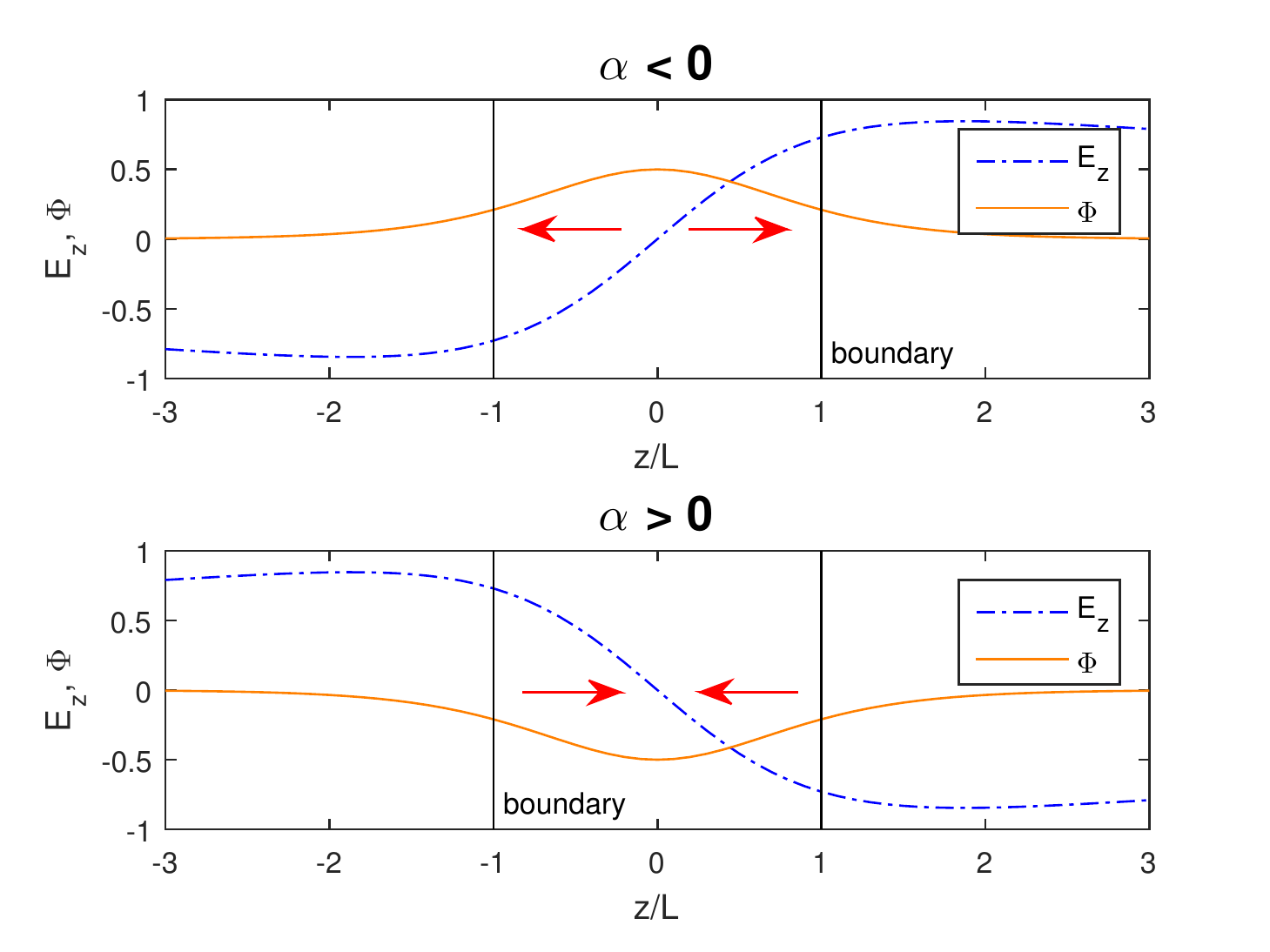}
\caption{The electric field and scalar potential; arrows represent the expected effect on ion motion for various $\alpha$.}
\label{electricFieldscalarPotential}
\end{figure}

Deformation of particle trajectories in the $(\kappa x, p_x)$ plane for $\alpha\ne 0$ can be explained by taking into account the relation between the $\kappa x$ and $z$ particle coordinates. Ions trapped near the neutral plane will stay in the $(\kappa x, p_x)$ plane near $\kappa x<1$; ions that escape from the neutral plane and reach large $|z|$ will move in the $(\kappa x, p_x)$ plane along trajectories that approach large $\kappa x$. This is illustrated by Fig. \ref{comparetrajalpha}, which shows that for $\alpha>0$ there are more ions trapped near the neutral plane while for $\alpha<0$, the electric field ejects ions far from the neutral plane. And, indeed, Fig. \ref{electricFieldscalarPotential} shows that $\alpha>0$ corresponds to the scalar potential trapping positively charged ions, whereas $\alpha<0$ corresponds to the scalar potential reflecting ions away from the current sheet.

\section{Particle $v_y$ velocity}
Charged particles moving along closed trajectories in the $(\kappa x, p_x)$ plane (see Fig. \ref{comparetrajalpha}) do not contribute to the total current density in the system\cite{Pellat&Schmidt79}. However, an electrostatic field $E_z$ causes a particle drift motion along the $y$-direction even with demagnetized particles. To estimate this effect, we consider the $v_y$ component of particle velocity, which is defined as (see Hamiltonian (\ref{dimensionlessHamiltonian}))
\begin{align}
  v_y = \frac{1}{2} z^2 - \kappa x
\end{align}
Velocity $v_y$ depends on the fast coordinate $z$ and thus oscillates significantly around some relatively slowly changing value. To obtain this averaged value depending only on slow variables, we calculate $\langle v_y \rangle$
\begin{align}
  \langle v_y \rangle = \oint v_y dt = \oint \frac{v_y}{p_z} dz \notag
\end{align}
We define the function
\begin{align}
	f_{v_y}(s) = \int_{\zeta_{\min}}^{\zeta_{\max}}  \frac{\frac{1}{2} y^2 - s}{\sqrt{1 - \big(s - \frac{1}{2}\zeta^2\big)^2}} d\zeta
	\label{fvys}
\end{align}
with the same $\zeta_{\min}$, $\zeta_{\max}$ as in Eq. (\ref{fs}).

\begin{figure}
\includegraphics[width=20pc]{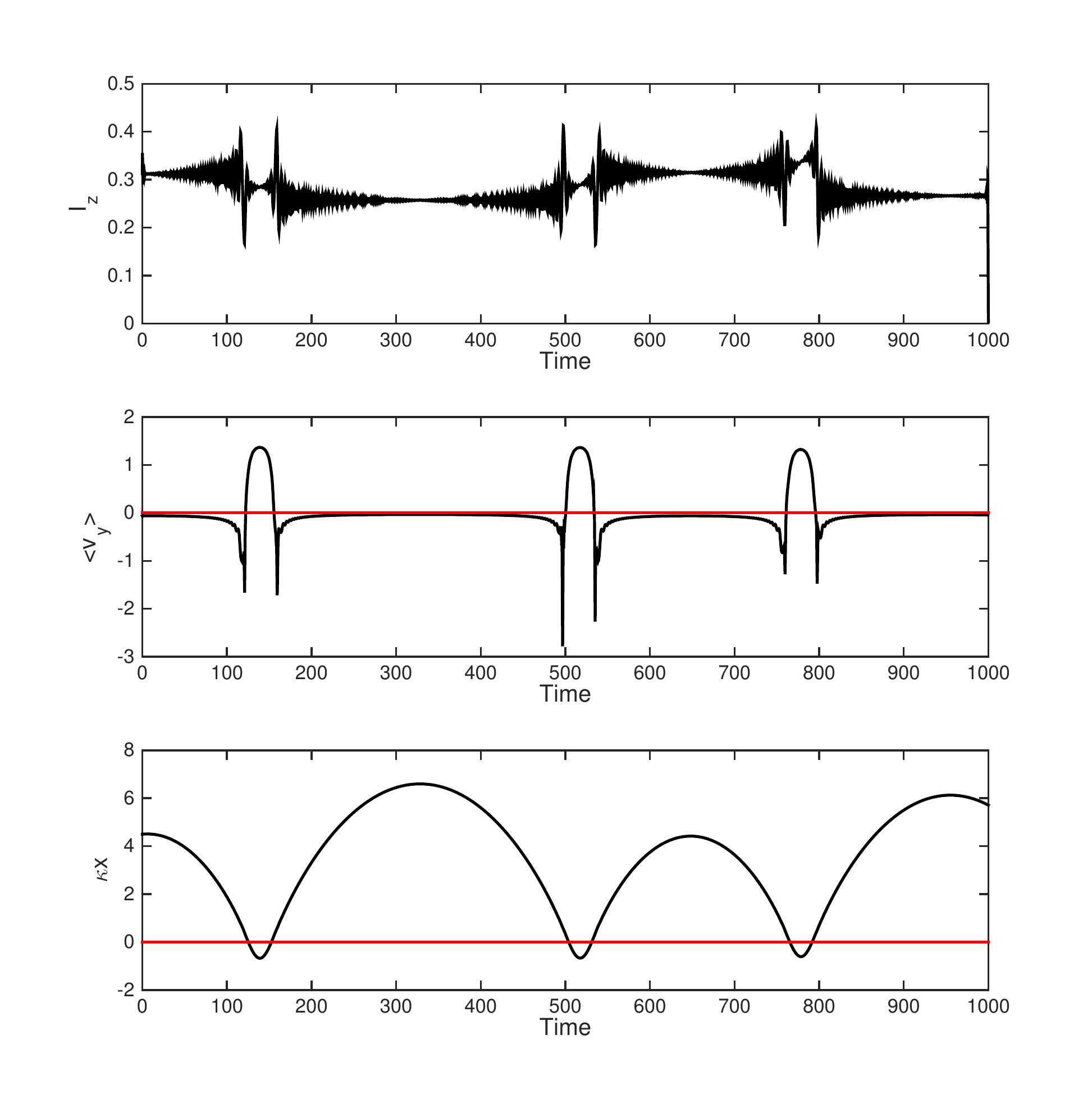}
\caption{Adiabatic invariant $I_z$, averaged particle velocity $\langle v_y\rangle$, and coordinate $\kappa x$ are plotted for a numerically integrated trajectory.}
\label{fvyskx}
\end{figure}

\begin{figure}
\includegraphics[width=20pc]{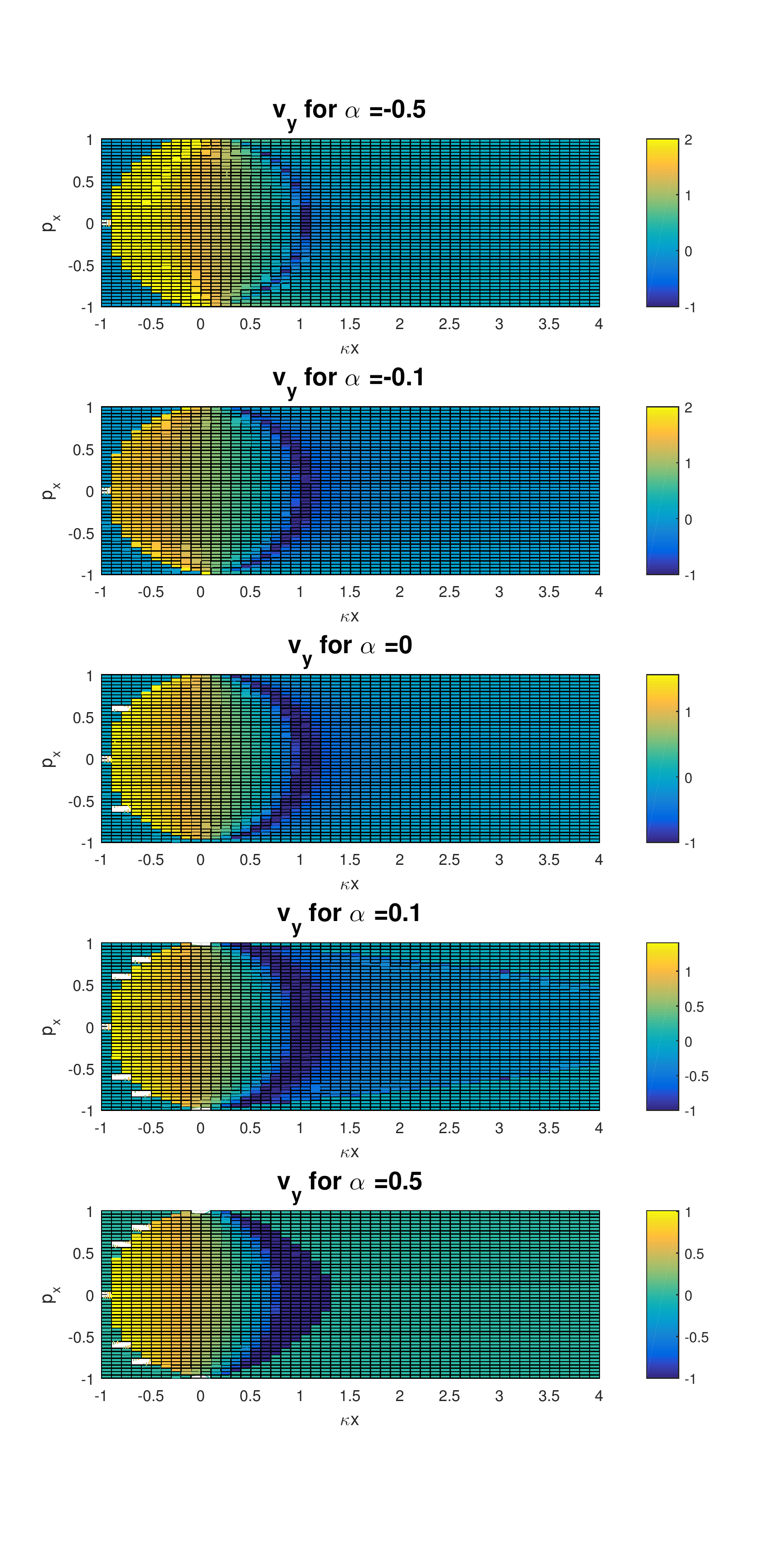}
\caption{Averaged particle velocity $\langle v_y\rangle$ for various $\alpha$ values.}
\label{comparealphasVy}
\end{figure}

For systems with $\alpha\ne 0$, the parameter $s$ in Eq. (\ref{fvys}) should be replaced by $\tilde{s}$ given by Eq. (\ref{syvariables_new}). We must be careful to note that the definition of the particle velocity does not depend on $\alpha$, thus rewriting $\langle v_y \rangle$ as
\begin{align}
  \langle v_y \rangle = \int_{\zeta_{\min}}^{\zeta_{\max}}  \frac{\frac{1}{2} \tilde{\zeta}^2 - s^*}{\sqrt{1 - \big(\tilde s - \frac{1}{2}\tilde{\zeta}^2\big)^2}} d\tilde{\zeta}
\end{align}
where $s^*=\kappa x/\sqrt{2\tilde{h}_z}$. Figure \ref{fvyskx} shows the profile of $\langle v_y \rangle$ along the particle trajectory obtained via numerical integration of the Hamiltonian equations (\ref{EoM}) for $\alpha=0$. The averaged velocity, $\langle v_y \rangle$, is positive when the particle oscillates across the neutral plane ($\kappa x<0$) and negative when the particle moves along magnetic field lines far from the neutral plane ($\kappa x>1$). Particle scattering characterized by $I_z$ jumps coincides with $\langle v_y \rangle$ zero crossings. Thus, a change of particle motion (from oscillations in one of two possible potential wells to oscillations inside a single potential well, see Fig. \ref{fastPotEnergy}) corresponds to $\langle v_y\rangle$ zero crossings. For particles with small enough $\kappa$, we can assume $I_z=const$ and consider a 2D map of $\langle v_y \rangle$ values in the $(\kappa x, p_x)$ plane. Figure \ref{comparealphasVy} shows that $\langle v_y \rangle$ becomes more negative as $\alpha$ becomes more positive, whereas $\langle v_y \rangle$ becomes more positive as $\alpha$ becomes more negative. This is an effect characterized by the average particle ${\bf E}\times {\bf B}$ drift, which, unfortunately, does not coincide with spacecraft observations. This doubly averaged (both over fast $z$-oscillations and slow periodical motion in the $(\kappa x, p_x)$ plane) velocity $v_y$ should be zero for $\alpha=0$ (see Ref. \onlinecite{Pellat&Schmidt79}), negative for $\alpha>0$, and positive for $\alpha<0$. However, particles moving along sufficiently stretched trajectories (along $\kappa x$) can escape from a realistic system with boundaries and do not have a chance to make a single closed trajectory in the $(\kappa x, p_x)$ plane. For such particles, the averaged $v_y$ is always positive for any $\alpha$. The important question is how $\alpha$ influences a number of such transient particles. Therefore, for an accurate estimation of the $E_z$ field effect on charged particle dynamics, we need to consider a magnetic field model that includes boundaries.

\section{Phase space volume occupied by transient particles}
To investigate how the number of transient particles depends on electrostatic fields, we need to modify our magnetic field model (\ref{analyticalModel}) to include current sheet boundaries where $B_x \to const$. In a constant magnetic field, particles move freely along magnetic field lines. Thus, particles reaching the current sheet boundaries with $p_x>0$ cannot return to the neutral plane and escape from the system. These particles are denoted as transient.
Because of the openness of their orbits, transient particles can carry significant current density (the segments of particle trajectories with $\langle v_y \rangle<0$ are much shorter than the segments with $\langle v_y \rangle>0$). Therefore, it is important to estimate the relative volume of the phase space occupied by transient particles.

\begin{figure}
\includegraphics[width=20pc]{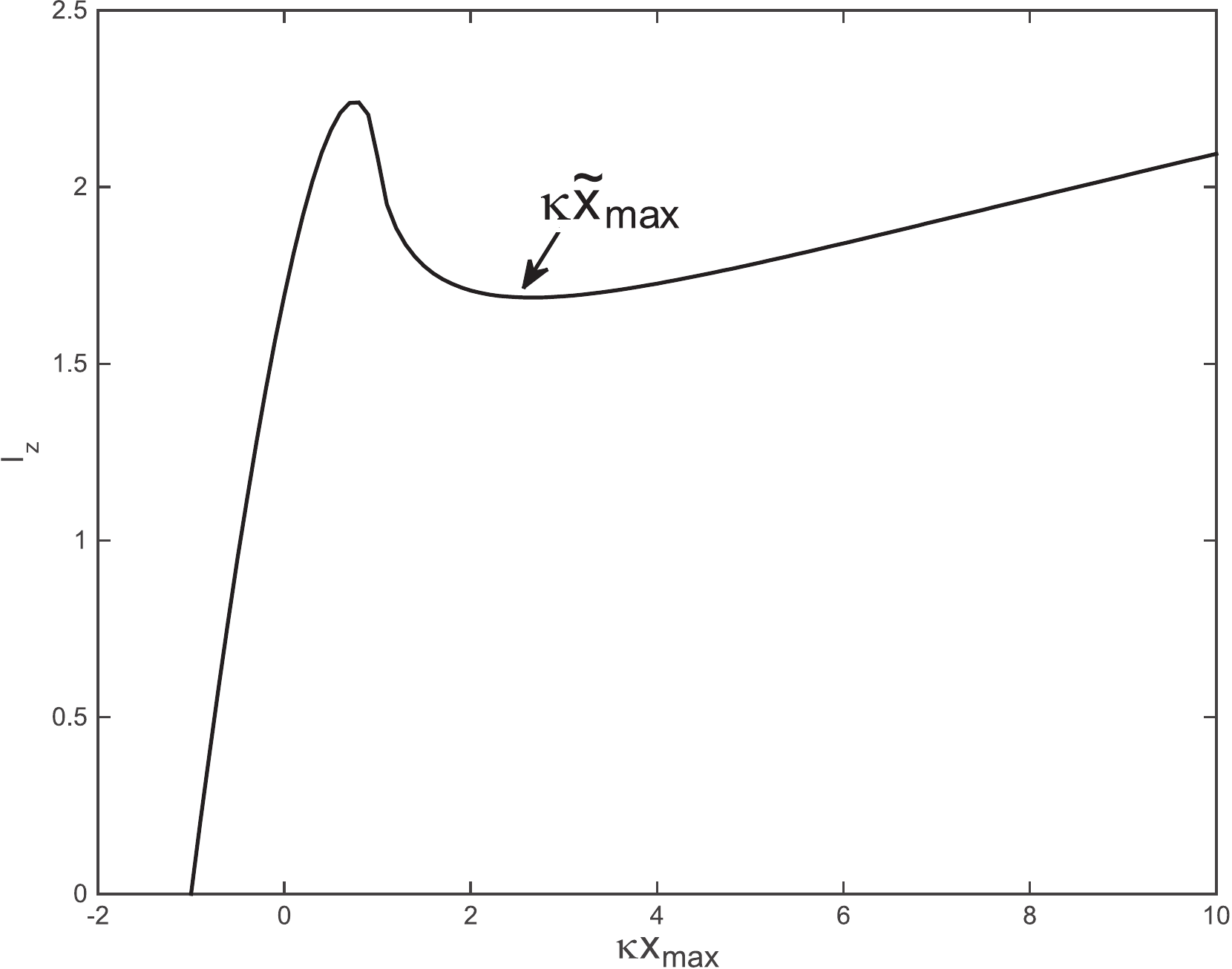}
\caption{Invariant $I_z$ as a function of $\kappa x_{max}$ for $\alpha < 0$}
\label{minIz}
\end{figure}

To separate transient from trapped particles, we introduce the model parameter $z_{\max}$, which denotes the position of the current sheet boundary. The Hamiltonian (\ref{dimensionlessHamiltonian}) contains the term $(\kappa x- z^2/2)^2$, whereas the total energy (Hamiltonian value) is conserved. Thus, for larger $z_{\max}$, the corresponding position of the current sheet boundary should be located around $\kappa x_{\max}\approx z_{\max}^2/2$. Particles with trajectories crossing $\kappa x_{\max}$ in the $(\kappa x, p_x)$ plane (see Fig. \ref{comparetrajalpha}) escape from the current sheet and can be considered as transient, whereas particles with trajectories crossing $p_x=0$ with $\kappa x<\kappa x_{\max}$ are trapped within the current sheet. To determine the area $S_{tran}$ occupied by transient particles in the $(\kappa x, p_x)$ plane, we calculate the total area occupied by particle trajectories $S_{total}$ for $\kappa x<\kappa x_{\max}$ and the area $S_{trapped}$ occupied by trapped particles. Area $S_{total}$ corresponds to the domain in the $(\kappa x, p_x)$ plane where
\begin{align}
  h_{z} = \mathcal H - \frac{1}{2}p_x^2 - \alpha \kappa x_{max} + \frac{1}{2}\alpha^2 > 0 \notag
\end{align}
The value of $S_{trapped}$ can be calculated as the area of the region surrounded by the ion trajectory touching $\kappa x_{\max}$  at $p_x=0$. This condition works well for $\alpha\geq 0$. However, for $\alpha < 0$, ions are reflected from the current sheet by the electric field. In this case, the trapped ion region's actual boundary can cross $p_x=0$ with $\kappa x<\kappa x_{\max}$. For a system with $\alpha < 0$, we find the actual $\kappa \tilde{x}_{max}$ to be equal to the maximum value of $\kappa x$ that corresponds with a trapped trajectory (see Fig. \ref{comparetrajalpha}).

The last trapped trajectory (closed trajectory with the largest $\kappa x$ at $p_x=0$) corresponds to some value of the adiabatic invariant $\tilde{I}_z$. This invariant equals $I_z$ calculated at $p_x=0$, $\kappa x=\kappa x_{\max}$ (for $\alpha\geq 0$) and at $\kappa \tilde{x}_{max}$ for $(\alpha<0)$, where $\kappa \tilde{x}_{max}$ corresponds to the position of $f(s)$ minimum (see Fig. \ref{minIz}). To calculate $\tilde{I}_z$, we use $\tilde{s}$ evaluated at
\begin{align}
  h_{z0} = h_z \Big|_{p_x=0} = \mathcal H - \alpha \kappa x_{max} + \frac{1}{2}\alpha^2 \notag
\end{align}
so all particles with $I_z>\tilde{I}_z$ are trapped.

\begin{figure}
\includegraphics[width=20pc]{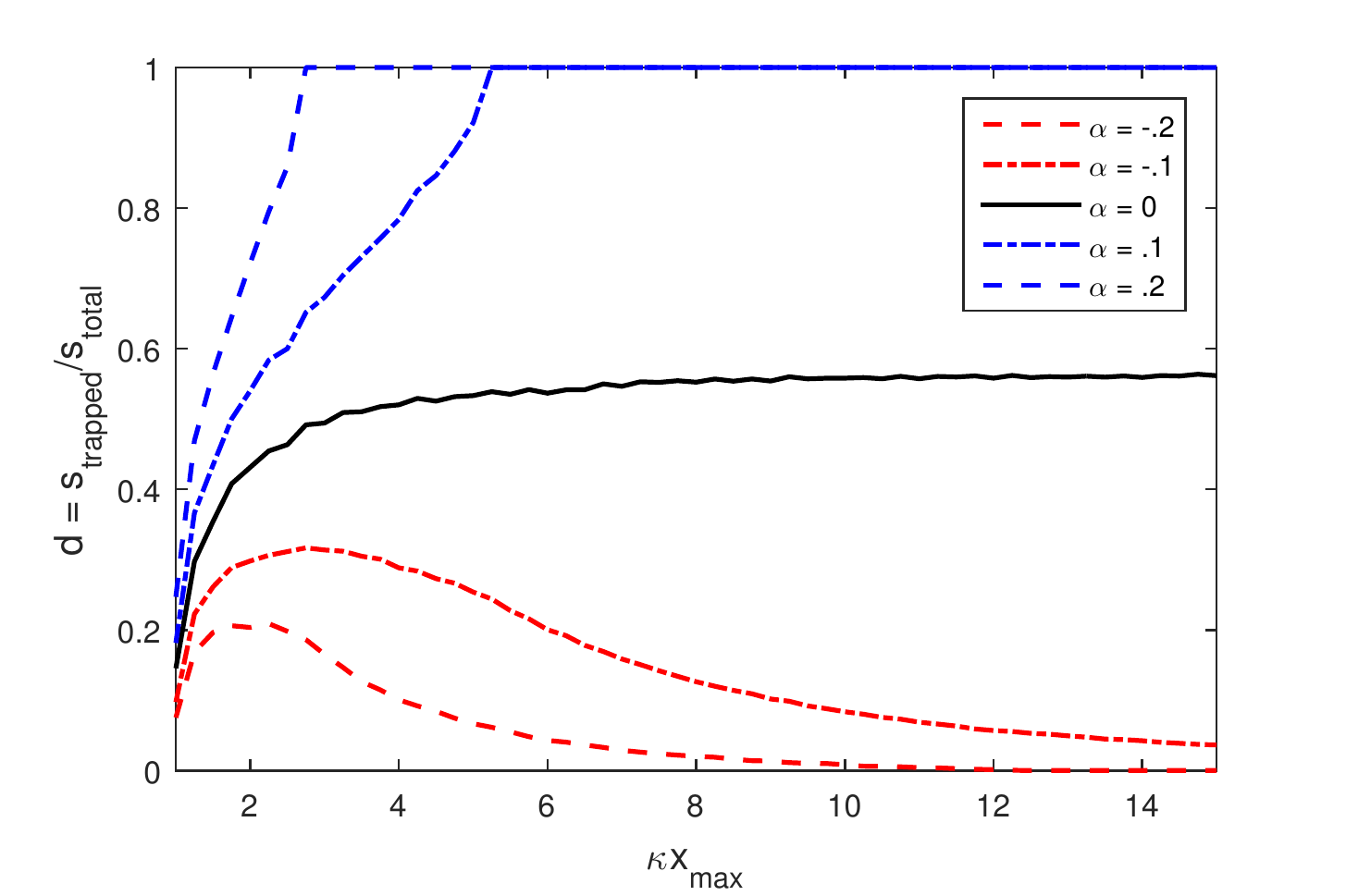}
\caption{$d(\kappa x_{\max})$ for various $\alpha$ values.}
\label{trappedRatio}
\end{figure}

We now define the relative area of transient particles as equal to $d_{trans}=1-d$ with $d=S_{trap}/S_{total}$. Figure \ref{trappedRatio} shows $d$ as a function of the current sheet boundary and the $\alpha$ value. For small $\kappa x_{max}$, an increase in $\kappa x_{max}$ results in a rapid increase in the relative number of trapped ions. The role of the current sheet width is apparent in Fig. \ref{trappedRatio}. The black line represents the result for zero electric field ($\alpha=0$). In this case, as $\kappa x_{max}$ increases, the trapped ions occupy a little more than half of the available area; transient ions occupy nearly the other half. For $\alpha > 0$ (blue), the electric field more quickly traps ions within the current sheet (see Fig. \ref{comparetrajalpha}) and thus $d$ rapidly approaches $1$. If $\kappa x_{max}$ and $\alpha$ are large enough, all ions can become trapped. For $\alpha < 0$ (red), ions are reflected from the current sheet and thus the area of trapped ions reaches a certain maximum value. Further $\kappa x_{\max}$ increases correspond with an $S_{total}$ increase, whereas $S_{trap}$ is conserved. Thus, $d(\kappa x_{\max})$ approaches zero for growing $\kappa x_{\max}$.

\section{Contribution to current density}
Knowing the relative phase space volume occupied by transient particles ($d_{trans} = 1 - d$ with $d=S_{trap}/S_{total}$), we can estimate their contribution to the current density. We begin with the Maxwell-Boltzman isotropic distribution of particle energy $\varepsilon$
\begin{align}
  f_{mb}(\varepsilon) = C n_0 \varepsilon^{1/2}\varepsilon_0^{-3/2} e^{-\varepsilon/\varepsilon_0}
\end{align}
where $C = 2/\sqrt{\pi}$ is the normalization constant, $n_0$ the density of particles, and $\varepsilon_0$ is the thermal (typical) energy. We pick typical parameter values for Earth's magnetotail current sheet\cite{Petrukovich15:ssr}: current sheet thickness $L \sim 1000$ km, $B_0 \sim 20$ nT, and $B_z \sim 2$ nT (i.e., $B_z/B_0 = 0.1$). The position of the current sheet boundary is modeled by the parameter $a=z_{\max}/L$. Therefore, for $\kappa \tilde{x}_{bound}$ we get
\begin{align}
  \kappa \tilde{x}_{bound} = \frac{a^2}{2} \frac{L}{\rho_0}
  = \frac{a^2}{2} \Bigg(\kappa(\varepsilon) \frac{B_0}{B_z} \Bigg)^2
\end{align}
where $\kappa$ depends on $\varepsilon$ as: $\kappa\approx 0.21\varepsilon^{-1/4}$ (we use the current sheet parameters listed above; energy $\varepsilon$ is in keV). Now, we can obtain $d(\varepsilon)$ and calculate a relative number of transient particles
\begin{align}
  N(a,\alpha) = \frac{n_{trans}}{n_{0}} = \frac{2}{\sqrt{\pi}}	\int_{w_c}^\infty d_{trans}(w) \sqrt{w} e^{-w} dw \label{eq:N}
\end{align}
where $w=\varepsilon/\varepsilon_0$ and $w_c$ corresponds with the critical value $\kappa_c=0.1$ (particles with $\kappa>\kappa_c$ cannot follow transient trajectories due to motion chaotization via scattering\cite{BZ89,Zelenyi13:UFN}). The transient particle velocity along the current density direction is approximately equal to the total particle velocity, i.e., when transient particles move along the $y$ direction in the neutral plane, almost all their energy is concentrated into $\langle v_y\rangle$ (see typical velocity distributions of transient particles in Refs. \onlinecite{Sitnov06, Zhou09, Artemyev&Zelenyi13}). Therefore, to estimate a transient particle current density, we can use an isotropic distribution and calculate particle flux
\begin{align}
  J(a,\alpha) = \frac{j_{trans}}{e n_{0}\nu_0} = \frac{2}{\sqrt{\pi}}	\int_{w_c}^\infty d_{trans}(w) w e^{-w} dw \label{eq:J}
\end{align}
with $\nu_0=\sqrt{2\varepsilon_0/m}$. The same properties, $N$ and $J$, are also calculated for a power-law distribution often observed in space plasma systems\cite{Livadiotis15}
\begin{align}
  f_{pl}(\varepsilon) = C n_0 \sqrt{\frac{\varepsilon}{\varepsilon_0}} \Big( 1 + \frac{\varepsilon}{k \varepsilon_0} \Big)^{-k-1} \label{eq:distPL}
\end{align}
with
\begin{align}
  C = \frac{2}{\sqrt{\pi}} \frac{\Gamma(k+1)}{k^{3/2} \Gamma(k-\frac{1}{2})}.
\end{align}
where $\Gamma$ is a Gamma function, and we use $k=3$. Substituting Eq. (\ref{eq:distPL}) into Eqs. (\ref{eq:N}, \ref{eq:J}), we obtain the  corresponding $N$ and $J$.

\begin{figure}[h]
\includegraphics[width=17pc]{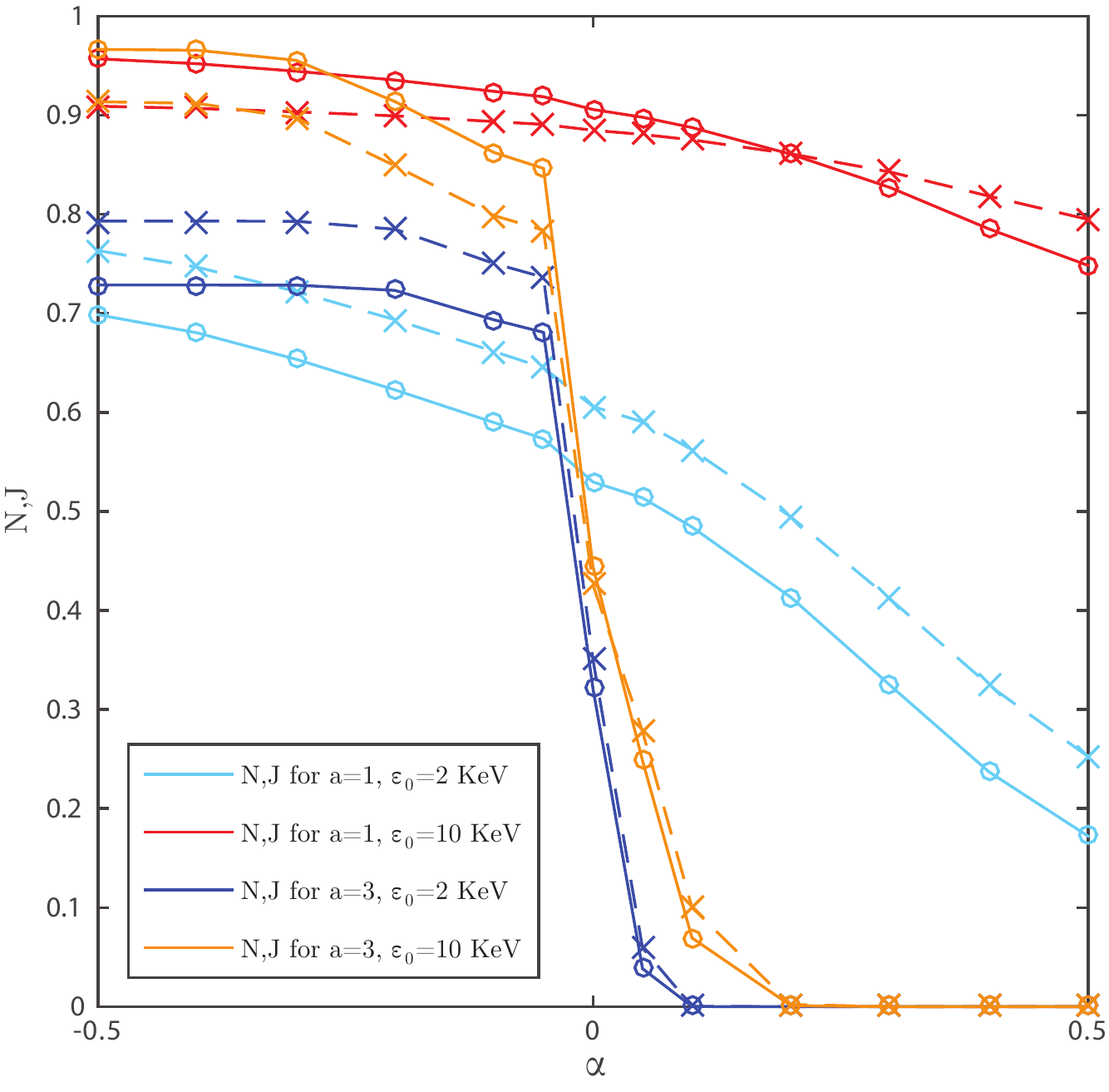}
\caption{$N$ (circles), $J$ (crosses) for various $\alpha$, $a$, and $\varepsilon_0$ (in keV) for a Maxwellian distribution.}
\label{maxwellianNJ}
\end{figure}

\begin{figure}[h]
\includegraphics[width=17pc]{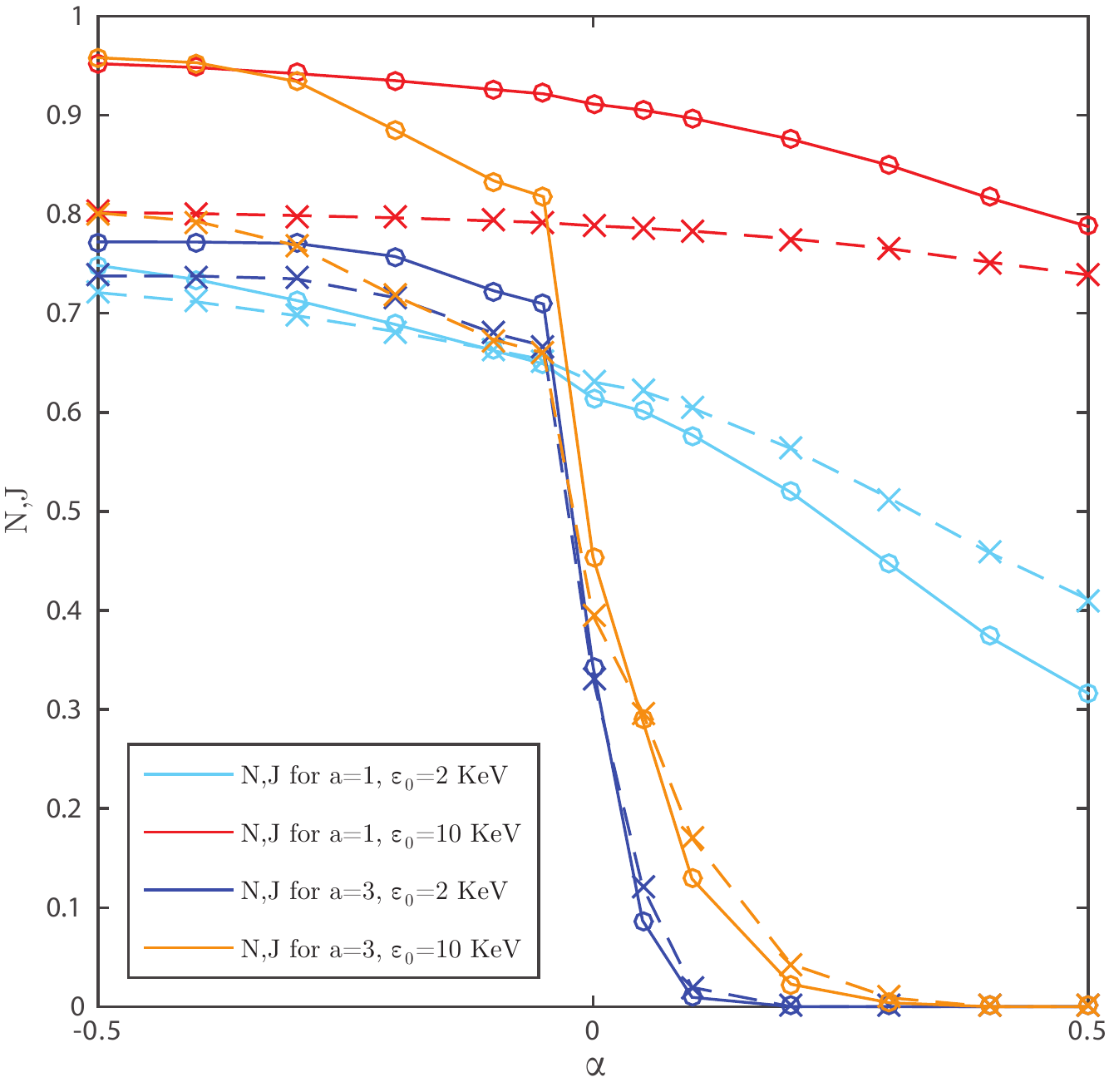}
\caption{$N$ (circles), $J$ (crosses) for various $\alpha$, $a$, and $\varepsilon_0$ (in keV) for a power law distribution}
\label{powerlawNJ}
\end{figure}

Figures \ref{maxwellianNJ}, \ref{powerlawNJ} show profiles $N(\alpha)$, $J(\alpha)$  for three $\varepsilon_0$ values and two $a$ values. There is very little difference between Maxwell and power-law distributions. The general trend is that positive values of $\alpha$ lead to ion trapping and thus a decrease in $N$ and $J$, whereas negative $\alpha$ corresponds to $d_{trans} = 1 - d$ growth (see Fig. \ref{trappedRatio}) and a corresponding increase in $N$ and $J$. Even a strong inward-directed electric field ($\alpha > 0$) cannot reduce the population of hot transient ions ($\varepsilon_0 = 10 \text{ keV}$) in a thin current sheet ($a = 1$) (red). While this electric field plays a more notable role in reducing cold ions ($\varepsilon_0 = 2 \text{ keV}$) in the thin configuration (light blue), the boundary distance plays the most significant part in determining the current density. Regardless of hot or cold ions, thick current sheet configurations ($a = 3$) combined with a small positive $\alpha\sim .1$ can result in total vanishing of the transient ion contribution to the current density.


We should mention that both density and current estimates of transient ions are performed for simplified Maxwell and power-law energy distributions. Moreover, we assume that the distributions are isotropic and thus have the same shape at the neutral plane $z=0$ and at some distance away from this plane. However, in more sophisticated self-consistent models the electric field would modify particle distributions (this effect is more important for electrons, see, e.g., Refs. \onlinecite{Artemyev13:angeo:double_spectra, Egedal13, Zenitani&Nagai16}). Therefore, our results provide a more qualitative picture of the $E_z$ role in transient particle dynamics.

\section{Discussion and conclusions}

The main estimates in this study are obtained with the assumption of $I_z$ conservation. Indeed, for small enough $\kappa$, jumps of the adiabatic invariant $I_z$ are small and of order $\sim \kappa$ (see Fig. \ref{adiabaticInvariantJumps} and Refs. \onlinecite{Neishtadt86, Cary86}). However, for long time intervals, these jumps result in $I_z$ destruction. Such long time intervals corresponds with many jumps and many periods of particle rotation in the $(\kappa x, p_x)$ plane. Each such period takes $1/\kappa$ time and randomly changes $I_z$ by $\sim \kappa$. Over time, the sum of these jumps should equal zero, and only diffusion of $I_z$ with variance ${\rm var}(\Delta I_z)\sim \kappa^2$ occurs \cite{BZ89, Zelenyi13:UFN}. Thus, to significantly change $I_z$, we need to consider the time interval of $1/\kappa^3$, which is much longer than the time scale of the transient particles' motion. Therefore, scattering and $I_z$ destruction cannot appreciably change transient particle dynamics ($\alpha < 0$). Trapped particle dynamics ($\alpha > 0$), on the other hand, can be influenced by this process; indeed, if the system contains a large population of trapped particles, the long term dynamics can result in particle ejection. In this study though, we focus on transient trajectories and leave the question of trapped particle dynamics for further investigation. However, we can estimate the effect of $E_z$ on the scattering rate, $\Delta I_z$, using the equations we derived.

Particles passing through the current sheet experience a jump in the adiabatic invariant with an amplitude $\Delta I_z \approx (2/\pi)\kappa p_x^*$, where $p_x^*$ is a solution of the equation $\tilde{s}=1$ (see, e.g., review \onlinecite{Zelenyi13:UFN} and references therein). The first equation in (\ref{syvariables_new}) for $\tilde{s}$ gives the following solution for $p_x^*$: $2h_z=(\kappa x)^2$. We substitute this solution into Eq. (\ref{newhz}) and obtain: $2\tilde{h_z}=(\sqrt{2h_z}-\alpha)^2=(\sqrt{1-(p_x^*)^2}-\alpha^2)$, where we use $2h_z=2\mathcal H -p_x^2$ and $2\mathcal H=1$. From the adiabatic invariant definition at $\tilde{s}=1$, $I_z=2(2\tilde{h}_z)^{3/4}f(1)$, we obtain $2\tilde{h}_z=(I_z/I_z^*)^{3/4}$ with $I_z^*=2f(1)$. Therefore, we finally find $p_x^*=\sqrt{1-(\alpha+(I_z/I_z^*)^{3/8})^2}$. For small $\alpha$ we get $p_x^*=\bar {p}_x^*-(I_z/I_z^*)^{3/8}\alpha/ \bar {p}_x^*$, where $\bar {p}_x^*=\sqrt{1-(I_z/I_z^*)^{3/4}}$ is $p_x^*$ for $\alpha=0$. This expression shows that positive $\alpha$ decreases $p_x^*$ as well as the adiabatic invariant jumps.

One of the most interesting effects shown in our study is the role that the electric field $E_z$ plays in variation of transient particle current density. Indeed, in a plasma system with magnetized ions and electrons, an $E_z$ field cannot influence the total (ion plus electron) current density, because the ${\bf E}\times{\bf B}$ drift of ions compensates for ${\bf E}\times{\bf B}$ drift of electrons. However, in a thin current sheet with a transient ion population, the electric field $E_z$ directly influences the number of transient particles. Transient particles carry the maximum possible current in hot plasma systems (when bulk velocity is smaller than thermal velocity, e.g., in planetary magnetotails) because their $\langle v_y \rangle$ velocity almost equals their thermal velocity. Thus, a decrease in transient particle density should significantly decrease the ion contribution to the total current density. This may explain  why spacecraft observations in Earth's magnetotail have shown electron currents to be much stronger than ion currents\cite{Runov06, Artemyev09:angeo,Vasko15:jgr:cs}. Similar results are found in numerical simulations\cite{Lin14:hybrid_code, Lu16:assymetry}.

An electric field $E_z$ can play an essential role in ion energization within thin current sheets located near the reconnection region. In analogy with electron surfing acceleration\cite{Hoshino05, Artemyev13:angeo:double_spectra, Zenitani&Nagai16}, ions trapped near the neutral plane can be transported large distances along the $x$-axis by a convection electric field $E_y$. This transport in the 2D system (with $\partial/\partial x\ne 0$) results in ion heating and acceleration \cite{Vainchtein05, Drake09}. For adiabatic particles with $\kappa\ll 1$, this effect can be considered in the future using the equations derived in our study and proposed in Ref. \onlinecite{Vainchtein05, Zelenyi13:UFN} for 2D current sheets without $E_z$.

In summary, we consider transient charged particle motion in the current sheet with a polarized electric field $E_z$. Using adiabatic invariants, we describe a
role of $E_z$ in macroscopic current sheet characteristics:
\begin{enumerate}
\item Depending on its direction, $E_z$ results in particle trapping (if $E_z$ points towards the neutral sheet) or particle expulsion (if $E_z$ points away from the neutral sheet) from the current sheet. This effect respectively reduces or increases the relative phase space volume occupied by transient particles.
\item Variation of the phase space volume of transient particles significantly influences the relative number of transient particles in the current sheet and the corresponding current density.
\item In a current sheet with a significant contribution of transient ions to the total current density (i.e., in thin current sheets), an electric field $E_z$ directed toward the neutral plane can reduce the total current density through particle trapping and a corresponding reduction of the transient ion population. This effect and a simple ${\bf E}\times{\bf B}$ drift result in the transformation of an ion-dominated current sheet to an electron-dominated current sheet, where the main current density is carried by electrons.
\end{enumerate}

\begin{acknowledgments}
We are thankful to J. Hohl,  H. Su, and C. Wong for the help with the manuscript editing. A. Artemyev is grateful to Dr. S. Lu (UCLA) for fruitful discussion of obtained results. We acknowledge NASA contract NAS5-02099. We would like to thank the following people specifically: C. W. Carlson and J. P. McFadden for use of ESA data, D.E. Larson for use of SST data, K. H. Glassmeier, U. Auster and W. Baumjohann for the use of FGM data provided under the lead of the Technical University of Braunschweig and with financial support through the German Ministry for Economy and Technology and the German Aerospace Center (DLR) under contract 50 OC 0302.
\end{acknowledgments}

\appendix

\section{Electric field estimates}
Direct measurements of the $E_z$ electric field component in the magnetotail are rather difficult to perform since the usual spacecraft orientation assumes that electric field antennas are located within the spacecraft plane $(x,y)$. However, $E_z$ can be derived from plasma and magnetic field measurements. We consider measurements of the THEMIS C spacecraft, which visited the Earth's magnetotail around $-16$ Earth radii downtail in 2009\cite{Angelopoulos08:ssr}. With an electrostatic analyzer\cite{McFadden08:THEMIS} and solid-state telescope \citep{Angelopoulos08:sst} onboard, THEMIS C provides ion and electron moments of distribution functions (density $n$, bulk velocity ${\bf u}$, thermal pressure ${\bf \textrm p}$); a fluxgate magnetometer provides magnetic field measurements\cite{Auster08:THEMIS}.

THEMIS C profiled the current sheet's structure as the sheet flapped (oscillated) across the spacecraft \cite{Runov05}. Ion bulk velocity measurements made along the current sheet normal, $u_z$, are assumed to be close to the current sheet flapping velocity. This assumption can be checked by comparing the measured $u_z$ with the magnetic field rate of variation $dB_x/dt$. A good correlation between $u_z$ and $dB_x/dt$ allows us to reconstruct the current sheet's spatial scale (thickness) and its corresponding current density $j_y\approx \mu_0 (dB_x/dt)/u_z$ (see, e.g., Ref. \onlinecite{Sergeev98}). We use three events that correspond to THEMIS C crossing the current sheet with a good correlation between $u_z$ and $dB_x/dt$ (all events are taken from statistics published in Ref. \onlinecite{Artemyev16:jgr:pressure}). Knowing the current density, $j_y\approx \mu_0 (dB_x/dt)/u_z$, within the current sheet, we can compare it with direct measurements of ion $j_{yi}=en_iu_{y}$ and electron $j_{ye}=-en_iu_{ye}$ current densities (assuming that the quasi-neutrality condition $n_i\approx n_e$ is satisfied). We select events with $j_{yi}+j_{ye}\approx j_y$; thus, both $j_{yi}$ and $j_{ye}$ are measured reliably. We use the magnetic field $B_x$ as a measure of distance from the current sheet's neutral plane $z=0$; in other words, we compare profiles of $j_y(B_x)$, $j_{yi}(B_x)$, and $j_{ye}(B_x)$.

For the selected current sheets we thus obtain spatial distributions of ion and electron currents: $j_{yi}(B_x)$, and $j_{ye}(B_x)$. Ions, being demagnetized in the magnetotail, can move along rather complex orbits and generate current density due to the effects of the openness of their orbits (transient ions \cite{Artemyev&Zelenyi13}). Therefore, it is rather complicated to decompose sources contributing to ion currents. In contrast, the motion of magnetized electrons is well described by drift theory; this allows the electron current to be decomposed into contributions from various drifts. In the magnetotail, which is filled by anisotropic electrons, the electron current density $j_{ye}$ consists of three parts \cite{BookSJB66}: 1) drift current due to electric field, $\sim -en_i({\bf E}\times{\bf B})/B^2$; 2) diamagnetic current $\sim B_x(\partial p_{\bot e}/\partial z)/B^2$; and 3) current due to the anisotropy of electron pressure $\sim (p_{\parallel e}-p_{\bot e})(\partial B_x/\partial z)/B^2$ (where $p_{\parallel e}$, $p_{\bot e}$ are components of the electric pressure tensor projected onto the magnetic field direction). Measurements of electron moments and the approximation $\partial/\partial z\approx u_z^{-1}(\partial/\partial t)$ allow us to estimate both diamagnetic and anisotropic currents\cite{Artemyev16:jgr:pressure}. Thus, the difference between $j_{ye}$ and these two estimates provide the profile of the current density $\Delta j_{ye}(B_x)$ generated by the electron cross-field drift in electric and magnetic fields.

Assuming that the current sheet is 2D (i.e., exists in $(x,z)$ plane) and approximating the absence of gradients along magnetic field lines (see discussion of this approximation in Ref. \onlinecite{Artemyev&Zelenyi13}), we obtain $E_z(B_x)\approx -\Delta j_{ye}(B_x)B_x/n_i(B_x)e$. The same estimates can be obtained from the ion current density (i.e., $E_z(B_x)\approx \Delta j_{yi}(B_x)B_x/n_i(B_x)e$, where $\Delta j_{yi}$ is due to ion drift motion). The difference $\Delta j_{yi}$, however, can only be estimated from analysis of ion distribution functions \cite{Zelenyi10GRL}, which is much more complicated than the analysis of electron fluid motion providing $\Delta j_{ye}(B_x)$. Thus, we use $\Delta j_{ye}(B_x)$, which gives us a profile of the electric field across the current sheet (see Fig. \ref{scheme}). Possible error in the electric field estimates has a magnitude of about $\Delta \phi/L$, where $\Delta \phi$ is a scalar potential drop along magnetic field lines and $L$ is the current sheet thickness (which can be estimated as the integral $\int u_zdt$). The maximum possible value of $\Delta \phi$ is about the electron temperature $p_{\bot e}/n_ie$ (see, e.g., Refs. \onlinecite{Egedal13, Artemyev&Zelenyi13}). We have checked that this maximum value $\Delta \phi/L$ does not exceed $30$\% of estimated $E_z$ for our three events.

\bibliographystyle{elsarticle-harv}


\end{document}